\documentclass[sigconf]{acmart}

\usepackage{pgfplots}
\usepackage{hyperref}

\usepackage{bm}
\usepackage{multirow}
\usepackage{makecell}
\usepackage{booktabs} 
\usepackage{multirow,bigdelim}
\usepackage{easybmat}
\usepackage{threeparttable}

\usepackage{pgfplotstable}
\pgfplotsset{compat=newest}
\usepackage{caption}
\usepackage{subcaption}
\usepackage{tikz}

\usepackage{mathtools}
\usepackage{stackengine}

\definecolor{dgreen}{HTML}{219601}
\definecolor{bblue}{HTML}{4F81BD}
\definecolor{rred}{HTML}{C0504D}
\definecolor{oorange}{HTML}{ffa435}
\definecolor{ggreen}{HTML}{9BBB59}
\definecolor{dred}{HTML}{5c0802}
\definecolor{bblue}{HTML}{15b9ff}
  
  \definecolor{forestgreen}{rgb}{0.13, 0.55, 0.13}
  
\newcommand*\hexbrace[2]{%
  \underset{#2}{\underbrace{\rule{#1}{0pt}}}}

  \newcommand{\drawBox}[3] 
  {
  \begin{tikzpicture} 
  \def\w{1.1} 
  \def\x{#1} 
  \def\xl{#1-#2} 
  \def\xu{#1+#2} 
  \filldraw[fill=gray!, draw=white] (0,0) rectangle (\x,0.2);
  \draw [white] (0,0) rectangle (\w,0.2); 
  \draw (\x,0.1) -- (\xu,0.1) -- (\xu,0.15) -- (\xu,0.05);
  \draw (\x,0.1) -- (\xl,0.1) -- (\xl,0.15) -- (\xl,0.05);
  \end{tikzpicture} 
  }

  \newcommand{\boxNumberConf}[3] 
  {\drawBox{#1}{#2} & #1 & #2} 
  
    \newcommand{\drawVBox}[1] 
  {
  \begin{tikzpicture} 
  \def\h{0.25} 
  \def \w{0.2} 
  \def\y{#1*\h} 
  \filldraw[fill=forestgreen, draw=forestgreen] (0,0) rectangle (\w,\y);
  \draw [white] (0,0) rectangle (0\w,\h); 
  \end{tikzpicture} 
  }
  
      \newcommand{\drawFiveVBoxes}[5] 
  {
  \drawVBox{#1} \hspace{-0.8em}\drawVBox{#2}\hspace{-0.8em} \drawVBox{#3}\hspace{-0.8em} \drawVBox{#4} \hspace{-0.8em}\drawVBox{#5}
  }
  
\usepackage{booktabs} 

\setcopyright{rightsretained}

\acmDOI{10.475/123_4}

\acmISBN{123-4567-24-567/08/06}

\acmConference[WOODSTOCK'97]{ACM Woodstock conference}{July 1997}{El
  Paso, Texas USA}
\acmYear{1997}
\copyrightyear{2016}

\acmArticle{4}
\acmPrice{15.00}

\begin{document}
\title{Decision Making of Maximizers and Satisficers Based on Collaborative Explanations}

\author{Ludovik Coba}
\orcid{0000-0003-1905-7472}
\affiliation{%
  \institution{Free University of Bozen}
  \streetaddress{Piazza Domenicani 3}
  \city{Bolzano}
  \country{Italy}
}
\email{lucoba@unibz.it}

\author{Markus Zanker}
\affiliation{%
  \institution{Free University of Bozen}
  \streetaddress{Piazza Domenicani 3}
  \city{Bolzano}
  \country{Italy}
}
\email{mzanker@unibz.it}

\author{Laurens Rook}
\affiliation{%
  \institution{TU Delft}
  \streetaddress{Jaffalaan 5}
  \city{Delft}
  \country{The Netherlands}}
\email{L.Rook@tudelft.nl}

\author{Panagiotis Symeonidis}
\affiliation{%
  \institution{Free University of Bozen}
  \streetaddress{Piazza Domenicani 3}
  \city{Bolzano}
  \country{Italy}
  }
\email{psymeonidis@unibz.it}

\begin{abstract}
Rating-based summary statistics are ubiquitous in e-commerce, and often are crucial components in personalized recommendation mechanisms. Largely left unexplored, however, is the issue to what extent the descriptives of rating distributions influence the decision making of online consumers. We conducted a conjoint experiment to explore how different summarizations of rating distributions (i.e., in the form of the number of ratings, mean, variance, skewness or the origin of the ratings) impact users' decision making. Results from over 200 participants indicate that users are primarily guided by the mean and the number of ratings and to a lesser degree by the variance, and the origin of a rating. We also looked into the maximizing behavioral tendencies of our participants, and found that in particular participants scoring high on the {\it Decision Difficulty} subscale displayed other sensitivities regarding the way in which rating distributions were summarized than others.
\end{abstract}



\maketitle

\section{Introduction}

Although ratings are no longer the sole source of user feedback as they have been in early works on collaborative filtering algorithms, they are still important cues to rank items according to users' presumed tastes and preferences. Moreover, in e-commerce in general rating summarizations are typically an important decision aid visualizing the aggregate opinions of a multitude of users. The mean rating value, or the total number of ratings, are common mechanisms to rank large item lists. In recommender systems research, summary statistics of the rating behavior of a user's nearest neighbors have been classified as collaborative explanation types \cite{Friedrich2011ASystems}; already Herlocker et al. \cite{Herlocker2000ExplainingRecommendations} identified them to  be a compelling way to explain the data behind recommendations. 
Largely left unexplored, however, is the issue to what extent the specific characteristics of rating distributions influence the choices of online consumers. We conducted a choice-based conjoint (CBC) experiment to explore how different summarizations of rating distributions (i.e., the total number of ratings, mean, variance, skewness or the origin of the ratings itself) impact users' decision making. In a pre-study that considered solely mean and number of ratings as attributes, we identified that users are more strongly guided by the mean rating value. It should be noted that this earlier study put to test representative attribute levels from the movie domain, where the average number of ratings per item is typically in the high three digit, or low four digit numbers. In that study, we noticed that, at levels with lower numbers of ratings, the relative importance of the number of ratings versus the mean value grows. Therefore, in this study, we sought to select a domain that is more representative for e-commerce in general, namely tourism, and extracted different attribute values from TripAdvisor rating data. Furthermore, it includes the origin of rating as a separate attribute -- i.e., if the rating summarization was based on all user ratings or just on those from users similar to the current one -- in order to quantify the strength of personalized collaborative explanations versus a justification based on all ratings.  

In addition, since decision making strategies vary from person to person \cite{Knijnenburg2011EachSystems}, we hypothesized that users that could be characterized as high on dispositional maximization would behave different from those high on dispositional satisficing in such as choice experiment. In particular, we predicted this to occur, if we followed the recommendation of \cite{nenkov2008short} to examine the three maximization dimensions separately.

Results from more than 200 participants indicate that, in general, users are primarily guided by the mean and the number of ratings, and to a lesser degree by the variance, as well as the origin of a rating. However, when looking into the maximizing behavioral tendencies of our participants, we clearly observe different sensitivities regarding the way in which rating distributions impact users' choice behavior. For instance, participants scoring low on {\it Decision Difficulty} -- which is a sub-scale of the commonly used Maximizing Scale \cite{nenkov2008short} -- considered mean and number of ratings with nearly similar weight, and also were considerably intrigued by ratings originating only from similar users, while those scoring high on this dimension were clearly selecting the choice with the higher mean more often.
The results of this study, therefore, provide clear indications about the degree of the potential {\it persuasiveness} \cite{yoo2012persuasive} of different representations of rating summaries -- i.e., framing them into the context of all users, or only similar users and their respective rating distribution characteristics. This leads us to discuss algorithmic tunings of matrix factorization algorithms in the final section of this work. \par

After outlining related work in Section~\ref{sec:rel_work}, in Section~\ref{sec:cbd_design} we will provide a detailed description of the choice-based conjoint methodology that was used in the present study. Section~\ref{sec:results} will serve to present obtained results, and in Section~\ref{sec:discussion}, finally, we hypothesize on the implications for recommendation algorithms and future research.

\section{Related work}
\label{sec:rel_work}

Explaining recommendations is a salient topic in the field of recommender systems, and has received considerable attention in the past years \cite{Tintarev2015ExplainingEvaluation,Nunes2017ASystems}. 

Herlocker et al.\cite{Herlocker2000ExplainingRecommendations} compared 21 different styles of explanations and demonstrated that the rating histograms were users' preferred mechanism to render the data behind the recommendations transparent. These user style explanations have proven to be popular also in many other studies \cite{Cosley2003,Bilgic2005ExplainingPromotion} ever since; also in the very recent one of Kouki et al.\cite{Kouki2017UserExplanations}, where user-based explanations and high mean rating values were identified to be the most popular styles.

The ``user'' style justifies recommendations by providing information on how similar users (neighborhood-based) interacted with the recommended item. The neighborhood is inferred from similar past behavior between users -- like clicking, buying or ratings actions. ``User'' style justifications are usually presented by a sentence like ``Similar users to you rated this item: $\ldots$"  \cite{Nunes2017ASystems}, followed by a rating summary statistics. 

It has been acknowledged in the literature on recommender systems that decision making strategies vary from person to person, and that people differ in the extent to which they search for "the best possible choice", or rather opt for "a decent choice considering the circumstances" \cite{Jugovac2018InvestigatingRecommendations,Knijnenburg2011EachSystems}. Inspired by the seminal work of Herbert Simon on the satisficing nature of human decision making \cite{simon1955behavioral}, Barry Schwartz and colleagues developed a theory and self-report scale to assess individual differences in a person's maximizing behavioral tendencies \cite{schwartz2002maximizing}. The authors distinguish between \textit{maximizers}, which are people tending to bargain to obtain the best solution for themselves, and \textit{satisficers}, which are people that -- like in Herbert Simon's essays -- tend to settle for a "decent enough" solution. Empirical research has revealed distinct behavioral responses for maximizers and satisficers. In general, people determined by high levels of maximization find it more difficult to cope with a large number of choices (so-called choice overload), take longer to make their choices, are less committed to their choices, display lower satisfaction with their choices, socially compare with others who seem to be better of, and/or even regret their choices \cite{dar2009maximization,iyengar2006doing,misuraca2013time,schwartz2002maximizing,sparks2012failing}. 

Unfortunately, the sparse work on dispositional differences between maximizing versus satisficing in the setting of recommender systems has failed to replicate these findings. That is, \cite{Knijnenburg2011EachSystems} reported an opposing response pattern for satisfaction with choices derived from non-personalized recommendations (i.e., maximizers appeared more rather than less satisfied with their choices than satisficers), whereas \cite{Jugovac2018InvestigatingRecommendations} even reported null effects in the presence of recommendations. One possible explanation for these inconsistencies may be that those studies failed to take into account the theoretical building blocks underlying maximizing and satisficing. It is true that a person's behavioral tendency towards maximization could be measured and analyzed as an aggregate measure per se, but it may pay-off to study differences in maximizing-satisficing at a higher level of granularity -- that is, by also focusing on the three sub-dimensions (alternative search, decision difficulty, and high standards) that make up maximization, separately \cite{nenkov2008short}. 
 
Conjoint analysis is a widely appreciated methodological tool from marketing and consumer research, which is particularly applicable to the study of user preferences and trade-offs in the decision making process \cite{Rao2014ChoiceAnalysis}. A vast literature documents the merits of conjoint analysis for the study of marketing-related preference problems, as has been continuously reviewed in articles and book (chapters), cf., \cite{agarwal2015interdisciplinary,green1990conjoint,hauser2004conjoint,rao2008developments}. The conjoint methodology has also successfully been employed in a wide range of areas beyond marketing and consumer research, including education, health, tourism, and human computer interaction. In the latter domain, for instance, Cho et al.\cite{Cho2015TheStyle} used conjoint analysis to investigate elders' preference over smart-phone application icons. The authors explored the dynamics of two attributes (degree of realism and level abstraction) one with four levels and one with two levels, and ran their user study with a modest total of 30 respondents. Intriguingly, Marriott even used conjoint analysis to design its hotel chains, which highlights the practical value of the method; \cite{wind1989courtyard} referred to in \cite{rao2008developments}.

In the field of recommender systems and online decision support, Zanker and Schoberegger \cite{Zanker2014AnSystems} employed a ranking-based conjoint experiment to understand the persuasive power of different explanation styles over the users' preferences. 
More recently, and of interest to the present discussion, Carbonell et al. \cite{Carbonell2018ChoosingPhysician} observed that users select physicians based on considerations of user-generated content such as ratings and comments rather than the official descriptions of the physicians' qualifications. The authors relied on a choice-based conjoint design to understand the features influencing user's choice, and suggested that future consideration of such attributes in recommender systems would improve the decision making process.

While these studies seem to offer first evidence for the existence of a persuasive effect of the descriptive characteristics of rating summarizations, to the best of our knowledge, no study has explored this issue in relation to individual differences in maximizing versus satisficing behavioral tendencies. The novelty of our work, in that respect, is twofold: First, the inclusion of maximizing vs. satisficing user characteristics to account for differential tendencies the processing of rating summarizations is novel. Second, and related, the incorporation of these behavioral measures responds to the recent call to action on conjoint analysis in \cite{agarwal2015interdisciplinary,rao2008developments} to increase the knowledge on the ways in which people choose reference points from a wider list, and to use these insights to develop better utility models \cite{agarwal2015interdisciplinary}. As such, the present research goes well beyond the mere provision of a best practice in order to quantify the perceived utility of the characteristics of different rating summarizations.

\section{Methodology}
\label{sec:cbd_design}
We conducted a user study in order to understand the trade-off mechanisms between confrontation with different origins of ratings and item's ratings profiles. Our analysis was based on the Choice-Based Conjoint (CBC) methodology, which is also denoted as the Discrete Choice Experiment by several authors \cite{louviere2010discrete}. CBC analysis is an excellent method for determination of the impact of product features among consumers. It is frequently used in marketing and consumer research to determine user preferences over a wide range of product or service attributes ~\cite{Kuhfeld2010DiscreteChoice}. 

In conjoint designs, products (a.k.a., {\it profiles}) are modeled by sets of categorical or quantitative \textit{attributes}, which can have different \textit{levels}, cf. \cite{rao2008developments}. The CBC experiment is typically designed such that participants repeatedly select a preferred profile from varying \textit{sets of choices}. This design feature nicely matches real-world settings when users are confronted with recommendation lists \cite{chung2012general}. 

In the remainder of this Section, we discuss in greater detail how we investigated the user's decision making on collaborative explanations. We will elaborate on the way in which we designed the profiles (i.e., the rating summarizations). Also, we will outline our experimental procedure, and present the materials and measures used in the present research. 


\subsection{Attribute selection}
The first step in building a conjoint design is to determine the attributes and their corresponding levels. Rating summarizations are usually presented as a frequency distribution on the class of discrete ratings values, preceded by the origin of the rating. We specifically observed five distinct attributes: \textit{origin of ratings}, \textit{number of ratings}, \textit{mean rating}, \textit{variance} and \textit{skewness} of the ratings. We used these five attributes to develop our stimuli (i.e., the profiles) of ratings summarization.

The \textit{origin of ratings} is commonly used in the explanation of  recommendations. Different origins are considered to differentially influence the decision making process (i.e., in terms of persuasiveness, effectiveness, etc.), and improve the overall experience of the user on the platform~\cite{Tintarev2015ExplainingEvaluation}.
The total {\it number of ratings} is often seen as a proxy for an item's popularity, and many well-known algorithms are implemented to recommend items that are frequently rated~\cite{Jannach2015}. 
Following the argument of ~\cite{deLanghe2016NavigatingRatings}, a high number of ratings with a slightly lower rating mean should be preferred over higher means based on a much lower total number of ratings. This leads us to the third attribute of this study, the {\it mean rating value}. 
It should be noted that \textit{variance} and \textit{skewness} can be interpreted as measures of disagreement or conflicting opinions among prior reviewers on the platform. Even though the item might have a high overall score, variations of the scores, should, therefore, discourage the users from interacting with the item.

\begin{table}
\caption{Attributes and attribute levels.}

\begin{tabular}{llc}
\toprule
Attribute & Level & Value \\
\midrule
\multirow{2}{*}{A1: Origin of ratings} & L1 & Similar Users\\
&L2& All users\\
\multirow{2}{*}{A2: Number of ratings} & L1& 20\\
&L2 & 70 \\

\multirow{2}{*}{A3: Mean rating} & L1& 3.7\\
&L2 & 4.3 \\

\multirow{2}{*}{A4: Variance} & L1& 0.7\\
&L2 & 1.3 \\

\multirow{2}{*}{A5: Skewness} & L1& -1.2\\
&L2 & -0.5 \\
\bottomrule

\end{tabular}
\label{table:attributes_levels}
\end{table}

\begin{figure*}
 \begin{subfigure}[t]{0.24\textwidth}
        \centering
        \includegraphics[width=4.4cm]{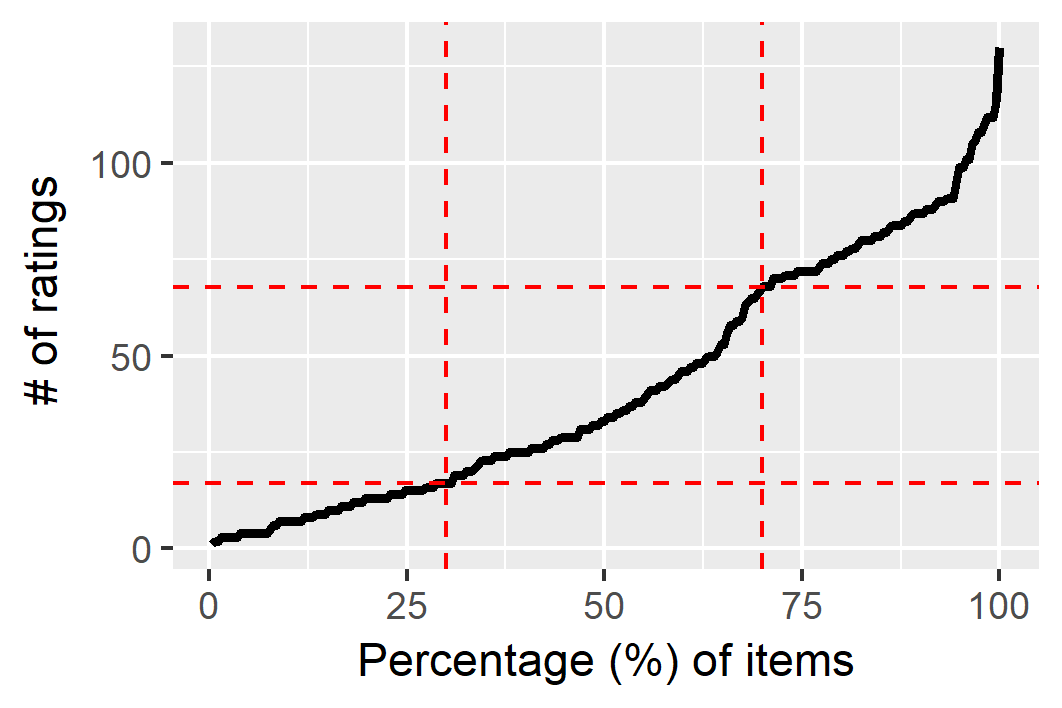}
        \caption{}
        \label{fig:nr_r}
    \end{subfigure}%
    ~ 
    \begin{subfigure}[t]{0.24\textwidth}
        \centering
        \includegraphics[width=4.4cm]{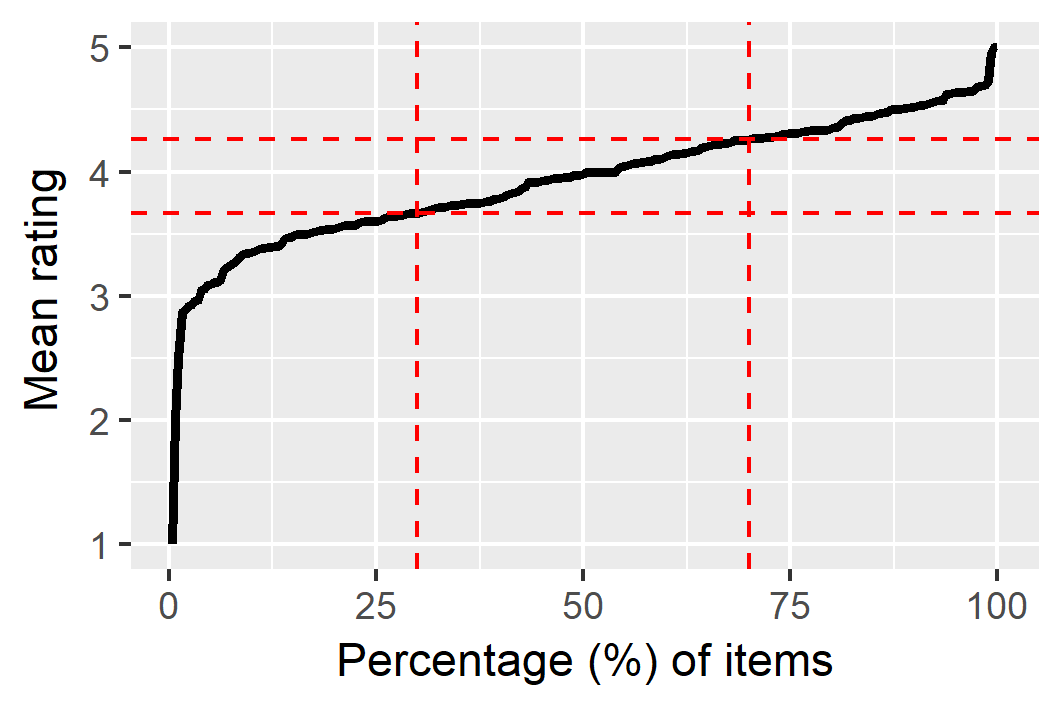}
        \caption{}
        \label{fig:mean}
    \end{subfigure}
     \begin{subfigure}[t]{0.24\textwidth}
        \centering
        \includegraphics[width=4.4cm]{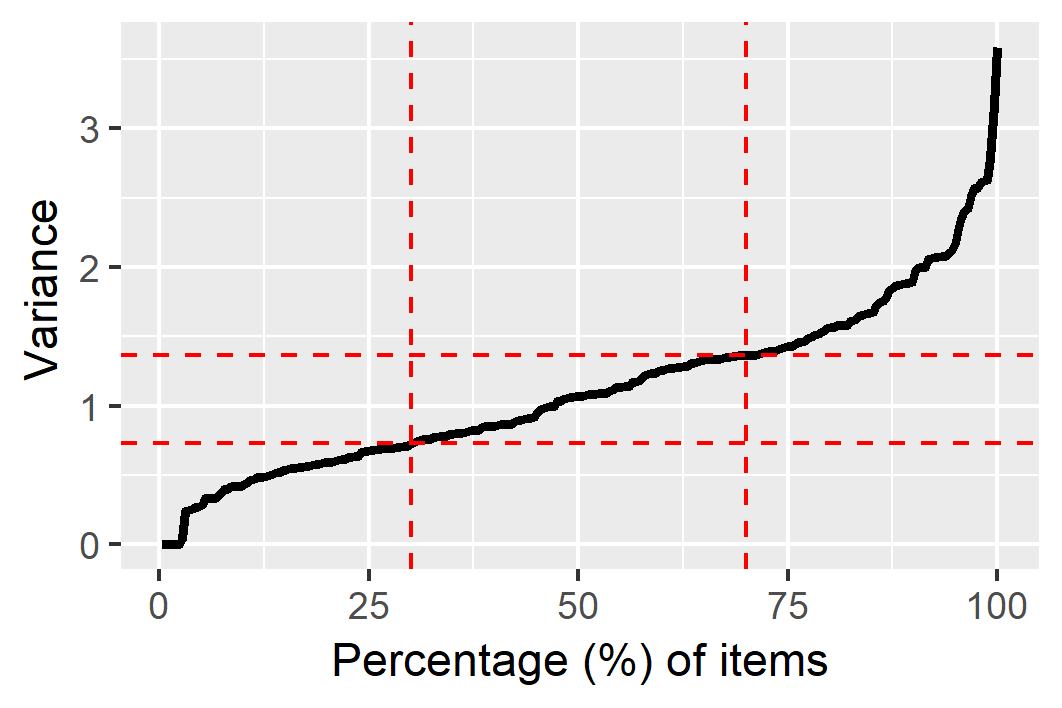}
        \caption{}
        \label{fig:var}
    \end{subfigure}%
    ~ 
    \begin{subfigure}[t]{0.24\textwidth}
        \centering
        \includegraphics[width=4.4cm]{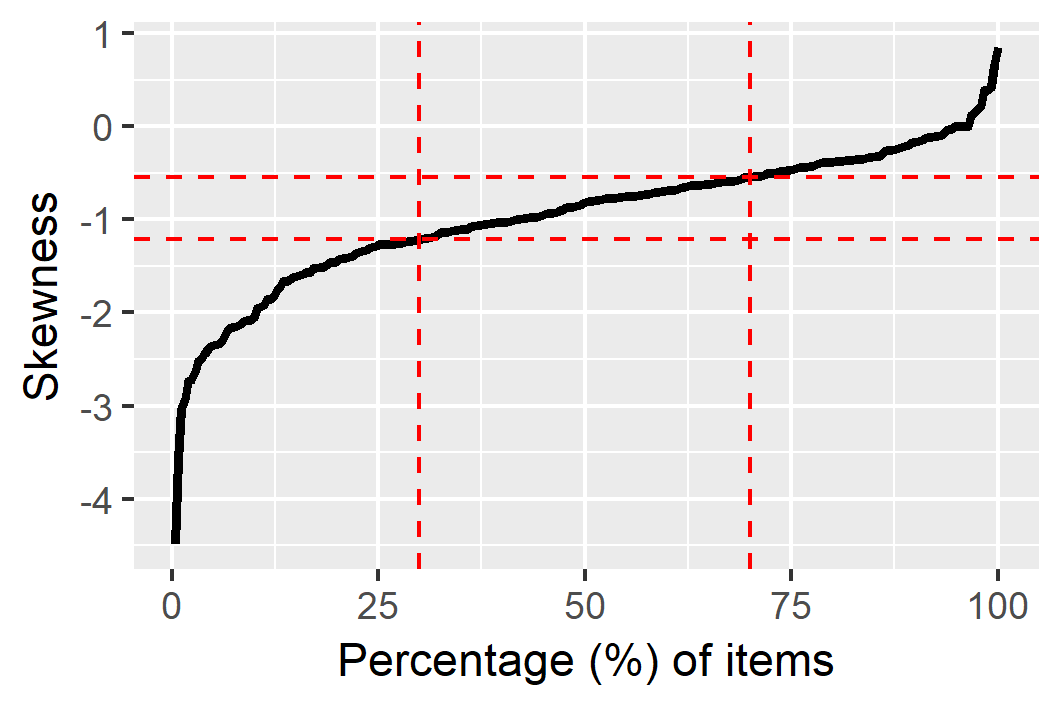}
        \caption{}
        \label{fig:skew}
    \end{subfigure}
    \caption{Rank distribution of items based on (a) number of ratings, (b) mean value, (c) variance and (d) skewness in the TripAdvisor dataset.}
    \label{fig:tp_analysis}
\end{figure*}

For the origin of ratings attribute we compared two levels: the personalized explanation (i.e., similar users), versus the unpersonalized explanation (i.e., all users). 
For this paper, we bootstrapped information about ratings from TripAdvisor's Web-crawled dataset obtained from \cite{Fuchs2012Multi-criteriaDomain}. This dataset contains all ratings of hotels from different destinations at the time of crawling. In order to bootstrap our choice experiments with realistic attribute levels, we filtered all hotels located in New York, which was the tourist destination with the highest number of reviews -- i.e., 11061 ratings on 258 hotels from a total 9597 named users. 
In order to determine the respective attribute levels, we analyzed the distribution of ratings per item in the TripAdvisor dataset (see Figure~\ref{fig:tp_analysis}). 
Figure~\ref{fig:nr_r} shows the rank distribution of the items based on the total number of ratings. 
The 30th and 70th percentiles of the number of ratings are 20 and 70, which we, henceforth, denote as the {\it Small} and {\it Large} levels of number of ratings. 
Next, Figure~\ref{fig:mean} depicts the rank distribution of the mean rating values. 
The 30th and 70th percentiles have rounded mean rating values of 3.7 and 4.3, respectively, which we transformed into the {\it Low} and {\it High} levels of our mean rating values. Figure~\ref{fig:var} shows the distribution of items based on the variance. Similar to the number of ratings and the mean value, the 30th and 70th percentiles yield variance values of 0.7 and 1.3, respectively, which became our {\it Low} and {\it High} levels of the variance. 
Finally, the skewness distribution is shown in Figure~\ref{fig:skew}. The 30th and 70th percentiles of the skewness distribution are -1.2 and -0.5, respectively, which we turned into our {\it Low} and {\it High} levels of skewness.

Table~\ref{table:attributes_levels} summarizes the selected attributes and the selected values for each level.

\subsection{Study design}

Conjoint choice experiments not only require a set of profiles; another requirement is a design outlining how these profiles are distributed into a number of choice sets, and presented to the respondents in the sample, cf. \cite{rao2008developments}. 

In the present study, the identified attribute levels allowed us to build a \textit{full-factorial design}~\cite{Zwerina1996ADesigns} that included all possible combinations of attributes and levels -- that is, a design, which consisted of 5 attributes $\times$ 2 levels each. This resulted in 32 different profiles that were put to the test (see Table~\ref{tab:profiles}).
Importantly, all profiles represented statistically feasible level combinations, while, for instance, a mean rating of 5 with a variance different from 0 would obviously be unfeasible.

\begin{table}
\footnotesize
\caption{Description of profiles. Orig. is the origin of ratings, Rt. refers to the number of ratings, Mean, Var., and Skew. correspond to the mean attribute, the variance attribute, and the skewness attribute, respectively. Ratings distribution shows, which percentage of the rated the item is T(Terrible), P(Poor), A(Average), V(Very Good) and E(Excellent).}
\begin{tabular}{c c c c c c c c c c c}
\toprule
\multirow{2}{*}{\makecell{Profile\\ID}}	&	\multirow{2}{*}{Orig.}	&	\multirow{2}{*}{Rt.}	&	\multirow{2}{*}{Mean}	&	\multirow{2}{*}{Var.}	&	\multirow{2}{*}{Skew.}	&	\multicolumn{5}{c}{Ratings Distribution}\\

\cline{7-11}
 
 
&&&&&&	T & P & A & V & E	\\
\midrule

1	&	S.U	&	20	&	3.7	&	0.7	&	-1.2	&	3\%	&	8\%	&	18\%	&	65\%	&	8\%	\\
2	&	S.U	&	20	&	3.7	&	0.7	&	-0.5	&	3\%	&	0\%	&	38\%	&	45\%	&	15\%	\\
3	&	S.U	&	20	&	3.7	&	1.3	&	-1.2	&	10\%	&	3\%	&	15\%	&	55\%	&	18\%	\\
4	&	S.U	&	20	&	3.7	&	1.3	&	-0.5	&	3\%	&	15\%	&	20\%	&	28\%	&	35\%	\\
5	&	S.U	&	20	&	4.3	&	0.7	&	-1.2	&	0\%	&	5\%	&	8\%	&	35\%	&	53\%	\\
6	&	S.U	&	20	&	4.3	&	0.7	&	-0.5	&	0\%	&	0\%	&	23\%	&	28\%	&	50\%	\\
7	&	S.U	&	20	&	4.3	&	1.3	&	-1.2	&	0\%	&	18\%	&	3\%	&	15\%	&	65\%	\\
8	&	S.U	&	20	&	4.3	&	1.3	&	-0.5	&	0\%	&	8\%	&	25\%	&	8\%	&	60\%	\\
9	&	S.U	&	70	&	3.7	&	0.7	&	-1.2	&	3\%	&	8\%	&	18\%	&	65\%	&	8\%	\\
10	&	S.U	&	70	&	3.7	&	0.7	&	-0.5	&	3\%	&	0\%	&	38\%	&	45\%	&	15\%	\\
11	&	S.U	&	70	&	3.7	&	1.3	&	-1.2	&	10\%	&	3\%	&	15\%	&	55\%	&	18\%	\\
12	&	S.U	&	70	&	3.7	&	1.3	&	-0.5	&	3\%	&	15\%	&	20\%	&	28\%	&	35\%	\\
13	&	S.U	&	70	&	4.3	&	0.7	&	-1.2	&	0\%	&	5\%	&	8\%	&	35\%	&	53\%	\\
14	&	S.U	&	70	&	4.3	&	0.7	&	-0.5	&	0\%	&	0\%	&	23\%	&	28\%	&	50\%	\\
15	&	S.U	&	70	&	4.3	&	1.3	&	-1.2	&	0\%	&	18\%	&	3\%	&	15\%	&	65\%	\\
16	&	S.U	&	70	&	4.3	&	1.3	&	-0.5	&	0\%	&	8\%	&	25\%	&	8\%	&	60\%	\\
17	&	ALL	&	20	&	3.7	&	0.7	&	-1.2	&	3\%	&	8\%	&	18\%	&	65\%	&	8\%	\\
18	&	ALL	&	20	&	3.7	&	0.7	&	-0.5	&	3\%	&	0\%	&	38\%	&	45\%	&	15\%	\\
19	&	ALL	&	20	&	3.7	&	1.3	&	-1.2	&	10\%	&	3\%	&	15\%	&	55\%	&	18\%	\\
20	&	ALL	&	20	&	3.7	&	1.3	&	-0.5	&	3\%	&	15\%	&	20\%	&	28\%	&	35\%	\\
21	&	ALL	&	20	&	4.3	&	0.7	&	-1.2	&	0\%	&	5\%	&	8\%	&	35\%	&	53\%	\\
22	&	ALL	&	20	&	4.3	&	0.7	&	-0.5	&	0\%	&	0\%	&	23\%	&	28\%	&	50\%	\\
23	&	ALL	&	20	&	4.3	&	1.3	&	-1.2	&	0\%	&	18\%	&	3\%	&	15\%	&	65\%	\\
24	&	ALL	&	20	&	4.3	&	1.3	&	-0.5	&	0\%	&	8\%	&	25\%	&	8\%	&	60\%	\\
25	&	ALL	&	70	&	3.7	&	0.7	&	-1.2	&	3\%	&	8\%	&	18\%	&	65\%	&	8\%	\\
26	&	ALL	&	70	&	3.7	&	0.7	&	-0.5	&	3\%	&	0\%	&	38\%	&	45\%	&	15\%	\\
27	&	ALL	&	70	&	3.7	&	1.3	&	-1.2	&	10\%	&	3\%	&	15\%	&	55\%	&	18\%	\\
28	&	ALL	&	70	&	3.7	&	1.3	&	-0.5	&	3\%	&	15\%	&	20\%	&	28\%	&	35\%	\\
29	&	ALL	&	70	&	4.3	&	0.7	&	-1.2	&	0\%	&	5\%	&	8\%	&	35\%	&	53\%	\\
30	&	ALL	&	70	&	4.3	&	0.7	&	-0.5	&	0\%	&	0\%	&	23\%	&	28\%	&	50\%	\\
31	&	ALL	&	70	&	4.3	&	1.3	&	-1.2	&	0\%	&	18\%	&	3\%	&	15\%	&	65\%	\\
32	&	ALL	&	70	&	4.3	&	1.3	&	-0.5	&	0\%	&	8\%	&	25\%	&	8\%	&	60\%	\\

\bottomrule
\end{tabular}
\label{tab:profiles}
\end{table}

Three principles needed to be respected in order to build the choice sets and to draw the most information on main effects and interactions: \textit{level balance}, \textit{orthogonality}, and \textit{minimal overlap}~\cite{Zwerina1996ADesigns}.
First, level balance requires attribute levels to appear with equal frequency in the different choice sets. Second, orthogonality ensures that main and interaction effects are uncorrelated -- something, which is achieved by having all attribute levels vary independently from each other. Overlap among attribute levels (i.e., identical attribute values for two or more profiles within the same choice set), finally, reduces the collected information.
We used the established D-efficiency metric to measure the statistical effectiveness of our design~\cite{Johnson2013ConstructingForce}:

 \begin{equation}
 D-\text{efficiency} = 100 \times \frac{1}{N \times |(\bm{X_C}' \bm{X_C})^{-1}|^{1/p}}
 \label{eq:d_eff}
 \end{equation}

Where $N$ is the number of observations in the design, as before, $p$ is the number of parameters, and $X_C$ is the standardized orthogonal contrast coding of the matrix $\bm{X}$ \cite{Kuhfeld2010}. In matrix $\bm{X}$, columns correspond to the levels of each attribute. Each $m$ rows of the matrix $\bm{X}$, Figure~\ref{fig:des_mat}, where a single row is a binary representation of a profile in a choice set ($\bm{X_n}$).

Coding is the process of replacing our design levels by the set of indicator or coded variables. To determine the efficiency of our design, we relied on standard orthogonal contrast coding as recommended by \cite{Zwerina1996ADesigns}. Please note that the sum of squares of the column in a standard orthogonal coding matrix is equal to the number of levels (e.g., if $X$ has two levels, the sum of squares of the columns of $X_C$ is 2). 
Thus, if $\bm{X}$ is orthogonal and balanced $\bm{X_C}' \bm{X_C} = N \bm{I}$, where $\bm{I}$ is a $p\times p$ identity matrix. In this case, the denominator terms in Formula~\ref{eq:d_eff} cancel each other, such that the efficiency is $100\%$.

\begin{figure}

\[ 
X = 
\begin{array}{c@{}c}
\left[
    \begin{BMAT}[5pt]{cc.cc.cc.cc.cc}{cc.cc}
      0& 1& 0& 1& 1& 0 & 0& 1& 0& 1\\
      1& 0& 1& 0& 0& 1 & 1& 0& 1& 0\\
      \vdots & \vdots & \vdots & \vdots & \vdots & \vdots  & \vdots & \vdots & \vdots & \vdots\\
      \vdots & \vdots & \vdots & \vdots & \vdots & \vdots  & \vdots & \vdots & \vdots & \vdots
    \end{BMAT}
\right] 
& 
\begin{array}{l}
  \\[-30mm] \rdelim\}{3}{1mm}[$X_1$]
\end{array} 
\\
\hspace{0pt} L_1 \hspace{4pt} L_2  \hspace{8pt} L_1 \hspace{4pt} L_2 \hspace{8pt} L_1 \hspace{4pt} L_2 \hspace{8pt} L_1 \hspace{4pt} L_2 \hspace{8pt} L_1 \hspace{4pt} L_2 
\\[-1ex]
\hexbrace{1cm}{A_1}\hexbrace{1cm}{A_2} \hexbrace{1cm}{A_3}\hexbrace{1cm}{A_4} \hexbrace{1cm}{A_5}
\end{array}
\]

\caption{Design matrix $X$, represented in Nonorthogonal Less-Than-Full-Rank Binary or Indicator Coding. }
\label{fig:des_mat}
\end{figure}

\begin{figure}
\centering
\includegraphics[scale=.60]{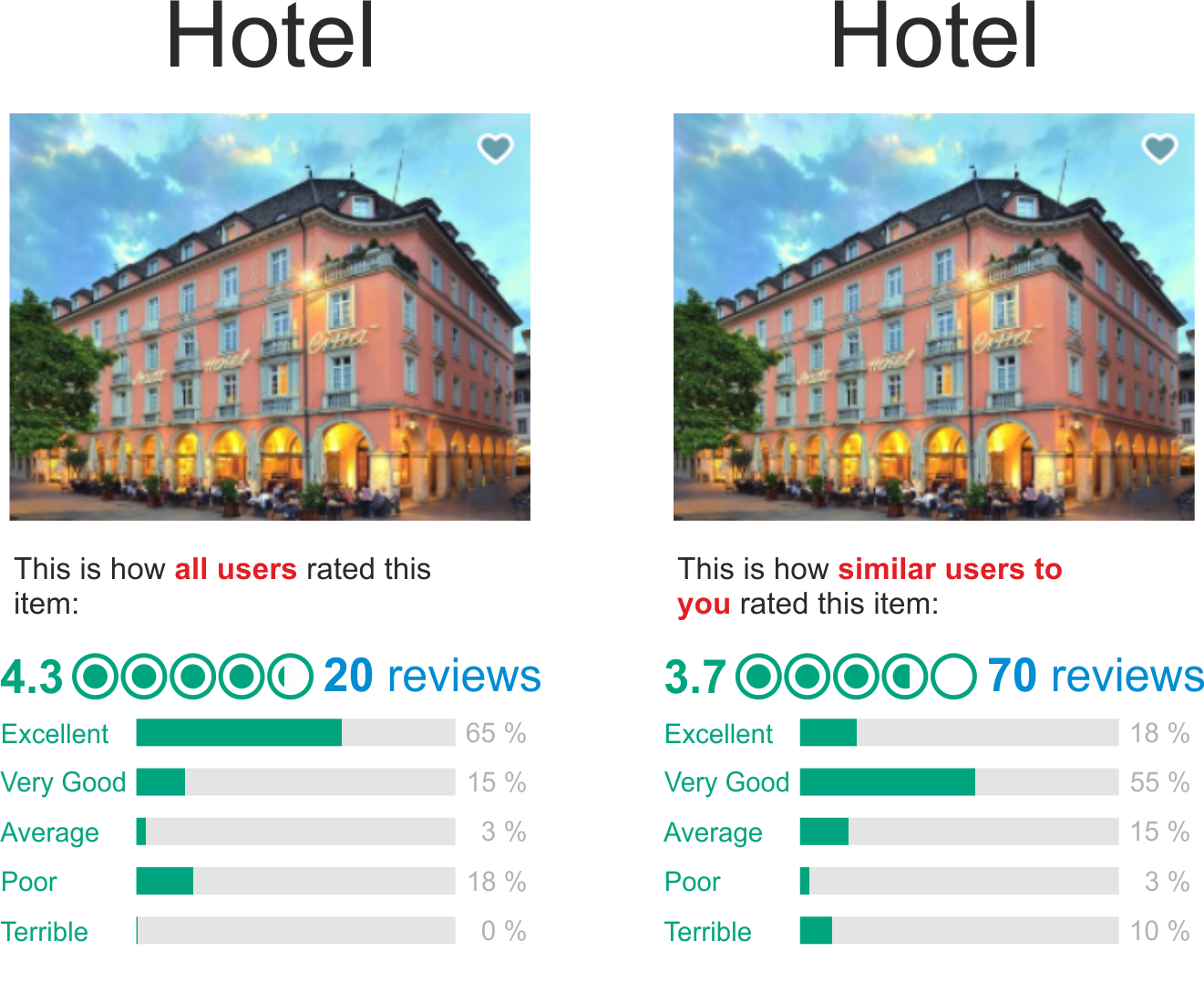}
\caption{An example snapshot of the binary choice between two rating summarizations based on different attribute levels. }
\label{fig:choice_set}
\end{figure}

We identified a CBC design consisting of $N=16$ choice sets with $m=2$ alternatives to be optimal due to attaining 100\% D-efficiency with minimal overlap, balanced frequency of levels, and orthogonality of effects. An example snapshot of a choice set is shown in Figure~\ref{fig:choice_set}.

\subsection{Statistical analysis}
In Figure~\ref{fig:choice_set} we depict an exemplary choice set consisting of two alternatives with different levels of the number of ratings and the mean rating value.
One of the basic assumptions underlying the assessment of users' choices is an additive utility model, assuming that the different attributes and characteristics of an item/profile will contribute, independently of each other, to the overall utility.

A respondent's preferences are modeled via a utility function $u(\bm{x_i})$~\cite{Zwerina1996ADesigns}, Formula~\ref{eq:utility}, representing how much the respondent likes a given item. 

\begin{equation}
\label{eq:utility}
u(\bm{x_i}) = \bm{x_i}\bm{\beta} + \epsilon
\end{equation}

where $\bm{x_i}$ is a vector characterizing a profile $i$, $\bm{\beta}$ is the vector with the unknown preferences for each attribute level, and $\epsilon$ is the residual error. The utility $u(\bm{x_i})$ of an item $\bm{x_i}$ is the sum of the partial utilities for each attribute. 

The most common approach in analyzing CBC-designs is the \textit{multinomial logistic regression} ~\cite{Hauber2016StatisticalForce,Rao2014ChoiceAnalysis,Zwerina1996ADesigns}, where -- given $N$ choice sets -- each consisting of $m$ profiles, the probability of choosing profile $i$ in the choice set $n$ is defined by Equation~\ref{eq:probability_of_choice}:

\begin{equation}
\label{eq:probability_of_choice}
P(\text{choice}_n = i) = \frac{e^{u(\bm{x_i}) }}{\sum_{j=1}^m e^{u(\bm{x_j}) }}
\end{equation}

Please note that the multinomial logistic regression is based on the assumption that the error $\epsilon$ is independent and identically distributed in a choice set.
Knowing the posterior probability, we use the multinomial logit to estimate the coefficients of vector $\bm{\beta}$ that maximize the likelihood of a profile to be chosen based on respondents' data.

\subsection{Personality scale}

Several scales exist to assess individual differences in maximizing versus satisficing behavioral tendencies, ranging from the 13-item Maximization Scale \cite{schwartz2002maximizing} to several shorter forms. In the present study, we used the shortened 6-item scale put forward by \cite{nenkov2008short}. Importantly, all these scales have in common that the behavioral tendency towards maximization -- even though it can be analyzed as a single, overall construct -- is better understood in terms of a three-dimensional disposition. Specifically, the psychometric qualities reveal the following sub-dimensions: \textit{alternative search}, \textit{decision quality}, and \textit{high standards}.

The following items in the shortened version by \cite{nenkov2008short} measured the sub-dimension \textit{alternative search}: "When I am in the car listening to the radio, I often check other stations to see if something better is playing, even if I am relatively satisfied with what I'm listening to", "No matter how satisfied I am with my job, it's only right for me to be on the lookout for better opportunities" (Cronbach's $\alpha$ = 0.30). The following items tapped into \textit{decision difficulty}: "I often find it difficult to shop for a gift for a friend", "Booking a hotel is really difficult. I'm always struggling to pick the best one"\footnote{Note, that we replaced {\it Renting a video [..]} in the original scale of \cite{nenkov2008short} with {\it Booking a hotel [..]}.} (Cronbach's $\alpha$ = 0.53). The sub-dimension \textit{high standards}, finally, was measured using: "No matter what I do, I have the highest standards for myself", and "I never settle for second best" (Cronbach's $\alpha$ = 0.74). The reliability measures for the first two sub-scales might appear low, however, they are perfectly within the ranges outlined by \cite{nenkov2008short}. Further, the overall scale was reliable; Cronbach's $\alpha$ = 0.53. Each of the items presented above was measured on a 7-point scale ranging from 1 ($completely$ $disagree$) to 7 ($completely$ $agree$).

\subsection{Study procedure}


Volunteers were invited per email to participate in an online user study on recommender systems and e-commerce. The interested volunteer was provided with a Web-link to the study, including a brief introduction into the research, and the guarantee that the data collection would safeguard anonymity. If the participant gave informed consent to have the data used for research purposes, she would be asked to fill out a short personality test (the shortened Maximization Scale described above). Next, the participant was asked to consider the following, hypothetical yet plausible, tourism-inspired decision making task:

\textit{``You need to make a choice between two hotels on a booking platform for your holiday stay. Both hotels are equally preferable to you with respect to cost, location, facilities, services, etc. Other users' ratings are aggregated and summarized by their number of ratings, the mean of their ratings, their distribution over the different rating values as well as by the origin of ratings (i.e. if the ratings were based on all users or from users similar to you).
Given the above, which of the two hotels below would you choose, when you were to solely consider the ratings for the two accommodations?''}

Following this introduction, the participant went through 16 choice tasks according to our design -- an exemplary choice set of which is presented in Figure~\ref{fig:choice_set}. The order of the choice tasks and the answer options (i.e., the profiles) were randomized for each respondent. Additional feedback on what characteristics of rating summaries guided their decision most, demographic information and general feedback on the questionnaire were included in the post-experimental part of the questionnaire.

\section{Results}
\label{sec:results}

\begin{table}
\caption{Summary of the respondents' demographics.}
\footnotesize

\begin{tabular}{lccccccc}
\toprule
Personal & \multicolumn{3}{l}{\multirow{2}{*}{Category}}&&&&\multirow{2}{*}{Total}	\\
feature\\
\midrule
Age & 18-24 & 25-30 & 31-40 & 40+ & & & \\
\# & 97 &  39 & 25 & 21 & & & 182\\
\% & 53\% & 21\% & 14\% & 12\%& & &100\%\\
\\
Gender & Female & Male & No answer&&&&\\
\# & 92 &  83 &7&&&& 182\\
\% & 51\% & 45\% &4\%&&&&100\%\\
\\
Country$^*$ & Italy & UK & Netherl. & Albania & Austria & Others \\
\# & 117 &  19 & 16 & 11 & 10&9 &182\\
\% & 64\% & 10\% & 9\% & 6\% & 5\% & 9\%&100\%\\
\bottomrule

\end{tabular}
\begin{minipage}[t]{0.42\textwidth}
\footnotesize
$^*$ Corresponds to the browser's geolocation from where the survey was accessed.
\end{minipage}
\label{tab:personal_data}
\end{table}

Between February and April 2018, \textit{215 subjects} from 12 countries participated in our study, where 182 completed the survey. In Table~\ref{tab:personal_data} we present the demographics of the participants in our sample. They were invited via the student mailing lists of management, economics and computer science faculties of our universities and per social media to participate in this study. No statistical differences were observed between the different demographic participant groups.

\begin{table}

\caption{Results of the multinomial logit.}

\begin{tabular}{lcc}
\toprule
Attribute	&	Level	&	Estimate ($\beta$)		\\
\midrule
\multirow{2}{*}{Origin}	&	Similar	&	0.37	(	0.05	) ***	\\
	&	All	&	-					\\
\multirow{2}{*}{\# ratings}		&	70	&	0.89	(	0.05	)	***	\\
	&	20	&	-					\\
\multirow{2}{*}{Mean}		&	4.3	&	1.18	(	0.05	)	***	\\
	&	3.7	&	-					\\
\multirow{2}{*}{Var.}		&	1.3	&	-0.18	(	0.05	)	***	\\
	&	0.7	&	-					\\
\multirow{2}{*}{Skew.}		&	-1.2	&	0.02	(	0.05	)		\\
	&	-0.5	&	-					\\
    \midrule
\multicolumn{3}{l}{Log-Likelihood:		-1484.8	}\\							\multicolumn{3}{l}{McFadden		$R^2$:	0.26	}\\
\multicolumn{3}{l}{Likelihood		ratio	test	: $X^2$=	1054.4 ***}\\
										
 \bottomrule
\end{tabular}

\begin{minipage}[t]{0.3\textwidth}
\footnotesize
\textit{Note:} ***	p<0.001;	** p<0.01;	* p<0.05. Dashes (-) are the baseline levels. The estimated coefficients are the change in log odds of choosing a particular mode rather than the baseline category. The values in parentheses are estimated standard errors.
\end{minipage}

\label{tab:res_all}
\end{table}
Since every participant had to complete 16 binary choice tasks, reported results are based on \textit{$182 \times 16 = 2912$} observed choices. For the entire sample of observations, we estimated the multinomial logit model underlying the CBC design, and report results in Table~\ref{tab:res_all}. The likelihood ratio test shows that the model was significant (McFadden's $R^2$ = 0.26, $X^2 = 1054.4, p<.001$). Presenting ratings only from similar users had a positive effect ($\beta$ = 0.37, p<.001) compared to the reference level -- displaying ratings of all users. As expected, a high number of ratings had a significant impact ($\beta$ = 0.89, p<.001), whereas the most significant influence derived from a high mean rating value ($\beta$ = 1.18, p<.001). Respondents also noticed the variance, and, as expected, were negatively influenced by the higher variance condition ($\beta$ = -0.18, p<.001). However, no significant influence was found for skewness ($\beta$ = 0.03, p>.05).
\begin{table}
\caption{Results of the multinomial logit for maximizers and satisficers$^\dagger$.}

\begin{tabular}{lccc}
\toprule
Attribute	&	Level	&	Maximizers ($\beta$) & Satisficers	($\beta$)	\\
\midrule
\multirow{2}{*}{Origin}	
&		Similar		&	0.34	(	0.07	)	 ***		&	0.39	(	0.07	)	 ***			\\
		&		All		&	-						&						-	\\
\multirow{2}{*}{\# ratings}			&		70		&	0.72	(	0.07	)	 ***		&	1.04	(	0.07	)	 ***		\\
		&		20		&	-						&						-	\\
\multirow{2}{*}{Mean}			&		4.3		&	1.14	(	0.07	)	 ***		&	1.23	(	0.07	)	 ***			\\
		&		3.7		&	-						&						-	\\
\multirow{2}{*}{Var.}			&		1.3		&	-0.18	(	0.07	)	 *		&	-0.17	(	0.07	)	 *		\\
		&		0.7		&	-						&						-	\\
\multirow{2}{*}{Skew.}			&		-1.2		&	-0.03	(	0.07	)			&	0.06	(	0.06	)				\\
		&		-0.5		&	-						&					-		\\

 \bottomrule

\end{tabular}

\begin{minipage}[t]{0.42\textwidth}
\footnotesize
 \textit{Note:} ***	p<0.001;	** p<0.01;	* p<0.05. Dashes (-) are the baseline levels. The estimated coefficients are the change in log odds of choosing a particular mode rather than the baseline category. The values in parentheses are estimated standard errors.
$^\dagger$ Comparison based on median split of the overall maximization score.
\end{minipage}

\label{tab:max_sat}
\end{table}
In addition, we analyzed results based on the participant's self-declared decision style based on the shortened Maximization Scale \cite{nenkov2008short}. In Table~\ref{tab:max_sat} we present the estimates from respondents, who scored high vs. respondents, who scored low on overall maximization. The multinomial logit model gives different partial utilities -- in particular for the number of ratings attribute -- of maximizers and satisficers that lead to higher probabilities of choice for rating summarizations with high rating numbers for satisficers.

\begin{table*}
\caption{Results of the multinomial logit for the three sub-scales$^\dagger$.}
 
\begin{tabular}{lccccccc}
\toprule
\multirow{2}{*}{Attribute}	&	\multirow{2}{*}{Level}	&	\multicolumn{2}{c}{Alternative search}	& \multicolumn{2}{c}{Decision difficulty}	& \multicolumn{2}{c}{High standards}\\
& & Low& High& Low & High & Low & High\\
\midrule

\multirow{2}{*}{Origin}		&		Similar		&	0.25	(	0.07	)	 ***		&	0.51	(	0.07	)	 ***	&	0.42	(	0.06	)	 ***	&	0.29	(	0.08	)	 ***	&	0.44	(	0.06	)	 ***	&	0.24	(	0.08	)	 **		\\
		&		All		&	-						&	-					&	-					&	-					&	-					&	-						\\
\multirow{2}{*}{\# ratings}			&		70		&	0.95	(	0.07	)	 ***		&	0.83	(	0.07	)	 ***	&	0.96	(	0.07	)	 ***	&	0.80	(	0.08	)	 ***	&	0.80	(	0.06	)	 ***	&	1.04	(	0.08	)	 ***		\\
		&		20		&	-						&	-					&	-					&	-					&	-					&	-						\\
\multirow{2}{*}{Mean}			&		4.3		&	1.29	(	0.07	)	 ***		&	1.05	(	0.07	)	 ***	&	1.09	(	0.07	)	 ***	&	1.31	(	0.08	)	 ***	&	1.12	(	0.06	)	 ***	&	1.30	(	0.08	)	 ***		\\
		&		3.7		&	-						&	-					&	-					&	-					&	-					&	-						\\
\multirow{2}{*}{Var.}			&		1.3		&	-0.11	(	0.07	)			&	-0.25	(	0.07	)	 ***	&	-0.14	(	0.06	)	 *	&	-0.24	(	0.08	)	 **	&	-0.20	(	0.06	)	 ***	&	-0.13	(	0.08	)			\\
		&		0.7		&	-						&	-					&	-					&	-					&	-					&	-						\\
\multirow{2}{*}{Skew.}			&		-1.2		&	0.03	(	0.06	)			&	0.01	(	0.07	)		&	0.02	(	0.06	)		&	0.02	(	0.07	)		&	0.02	(	0.06	)		&	0.02	(	0.07	)			\\
		&		-0.5		&	-						&	-					&	-					&	-					&	-					&	-						\\

 \bottomrule

\end{tabular}

\begin{minipage}[t]{0.91\textwidth}
\footnotesize
 \textit{Note:} ***	p<0.001;	** p<0.01;	* p<0.05. Dashes (-) are the baseline levels. The estimated coefficients are the change in log odds of choosing a particular mode rather than the baseline category. The values in parentheses are estimated standard errors.
$^\dagger$ Comparison based on median split of the respective maximization sub-scale.
\end{minipage}

\label{tab:subscales}
\end{table*}

However, following the recommendation of \cite{nenkov2008short}, we also looked into the details of the three components of the Maximization Scale. Table~\ref{tab:subscales} reports the separate results for a median split of each of the three sub-dimensions. Cheek and Schwarz \cite{cheek2016meaning} summarize that {\it Alternative Search} and {\it Decision Difficulty} both measure core components of the negative aspects of maximizing behavior that predict regret and dissatisfaction with life, as well as depression. In contrast, people scoring high on {\it High Standards} do not necessarily need to exhibit a maximizing behavior in their online decision making, but can also act as satisficers. According to \cite{cheek2016meaning} this is a potential reason for the inconclusive results of several studies that did not analyze the subscales but solely the overall maximization score; see also \cite{nenkov2008short} for the same point. 
Since, in our choice tasks, participants had no particular possibility to search for alternatives, but rather had to select one of the two options given the associated rating information, we were particularly interested in respondents experiencing {\it Decision Difficulty}. We do observe in Table~\ref{tab:subscales} that participants experiencing decision difficulty had a tendency to strongly rely on the higher mean ($\beta$ = 1.31, p<.001), and to avoid a high variance ($\beta$ = -0.25, p<.01) of rating data. They would also be less likely to select the alternative with the higher number of ratings, had they to accept a lower mean value in return. In contrast, respondents that scored low, seemed to be nearly equally likely according to their log odds to choose the high mean or the high number of ratings alternative -- that is, they could more confidently trade-in different attribute levels against each other in their decision strategy.

\begin{figure}
\centering
 \includegraphics[scale=0.7]{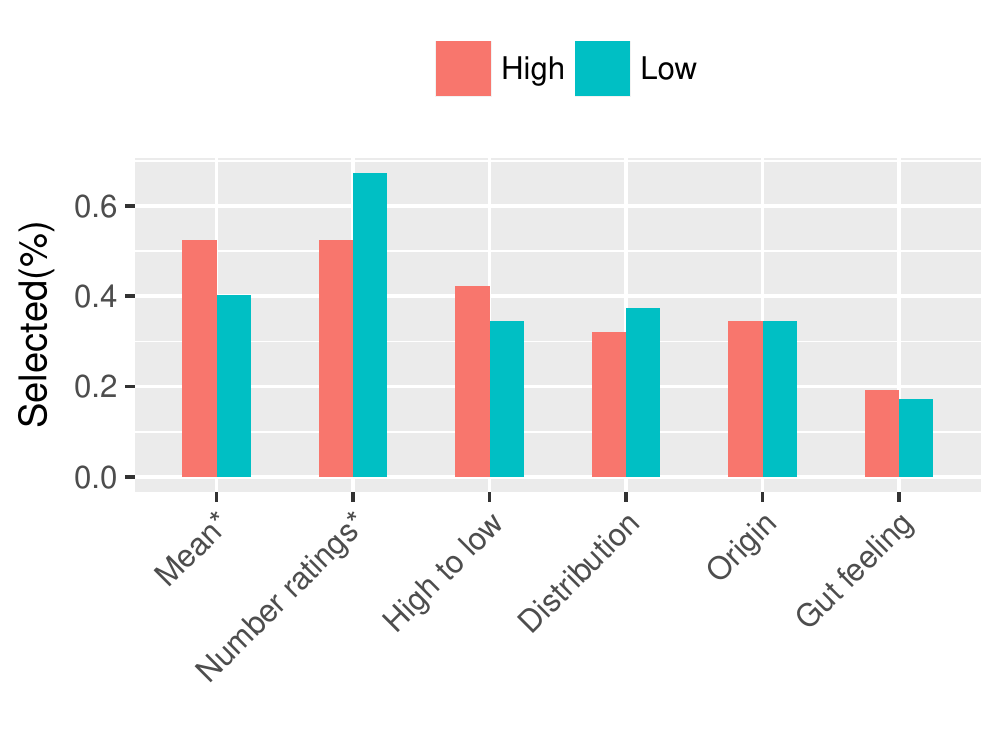}
\caption{Attributes that guided users' choices, median split of decision difficulty subscale. \textit{Note:} * significant, p<.05 one-sided.}
\label{fig/attributes}
\end{figure}

In the post-experimental part of the survey, we asked participants to self-report on the attributes they, retrospectively, thought to have guided their decision making behavior. This allowed us to test if participants were actually aware of their differences in choice behavior. Figure \ref{fig/attributes} plots the users' responses, separated for high and low scores on the Decision Difficulty subscale. The observation from the multinomial logit that participants scoring high on Decision Difficulty were more likely to go for high mean values and disregarded a high number of ratings was hereby clearly confirmed.

\section{Discussion and future work}

\label{sec:discussion}
\subsection{General considerations}


Results from our CBC experiment give us important clues how users value the five characteristics of rating summarizations (cmp. Table \ref{tab:res_all}). Marketing research has shown that consumers are strongly guided by online reviews, and that the mean rating value is interpreted as an indicator for the quality of a product \cite{duan2008online}. Also in our study, participants followed this quality hypothesis, where the mean characteristic is attributed the highest log odds from a multinomial logit. However, results also indicate that users display considerable sensitivity towards a larger number of ratings, in particular when they are solely double digit numbers, since they communicate a higher level of reliability and trustworthiness as mentioned by \cite{deLanghe2016NavigatingRatings}. Variance of ratings signifies disagreement among users and showed, as expected, a moderate negative effect on the choice probability of items. Higher skewness is in turn not really noticed to have an effect on the decision making of our participants. As an additional novel contribution, we also put the origin of ratings, i.e. summarizing all ratings vs. only a personalized subset of the users' nearest neighbors, as a justification for the presentation of an item to test. We observed that personalized rating summarizations possess a moderate positive effect on the probability of choice. However, since the number of ratings only from similar users needs to be obviously less than the total number of ratings, the negative effect of a lower number of ratings would invert the overall direction of the effect -- at least for the double digit rating numbers we studied here. Future work has therefore to determine the break-even point at which rating numbers presenting personalized rating histograms outperform unpersonalized rating summarizations.\par
These general results are therefore in line with prior research on the effects of potential decision biases \cite{chen2013human} that can be either purposefully exploited to develop more persuasive systems \cite{yoo2012persuasive} or explicitly neutralized, as has been, for instance, proposed by Teppan \& Felfernig\cite{teppan2012minimization}.

\subsection{Algorithmic considerations}

In line with the idea of tuning available algorithms to increase the actual probability of choice of presented recommendations Abdullahi \& Nasraoui \cite{Abdollahi2017UsingFactorization}, for instance, recently introduced an approach, denoted explainable matrix factorization.
They suggest that an item would be highly explainable, when having a high average rating in the neighborhood. In their line of argument users would thus benefit to a larger extent, if algorithms would take the presumed perception of explanations into consideration. They extend the matrix factorization (MF) loss function with a soft constraint that considers the perceived utility of each user based on the average mean of the ratings in a user's neighborhood.

Given our empirical findings, we can, however, propose that the utility of displayed collaborative explanations would not only depend on the mean of the ratings, but also on the total number of ratings and their variance. Thus, we can derive a multi-attribute utility of an item $j$ for user $i$ as follows: 

\begin{equation}
\label{eq:formula}
u_{ij} = \gamma_{i_{\#Rt}} \times \#Rt_j + \gamma_{i_{Mean}} \times Mean_j + \gamma_{i_{Var}} \times Var_j 
\end{equation}

Where $\gamma_{i_{\#Rt}}$,  $\gamma_{i_{Mean}}$ and $\gamma_{i_{Var}}$ are the parameter estimates for user $i$ that could also take differences in presumed decision styles into account. While, $\#Rt_j$ $Mean_j$ and  $Var_j$ are correspondingly the number of ratings, the mean, and the variance of item $j$. Please note, that due to this additive formula for utility weights, low partial utility values on one attribute can be compensated by higher partial utilities on another. Items scoring higher on such a utility function, should thus have higher odds to be included in actual recommendations under the condition of a similar matching score in terms of relevance for a particular user. Furthermore, we need to disclaim here, that additional more fine grained sensitivity results are needed to understand how the trade-off function between different characteristics looks like in order to determine the individualized attribute weights $\gamma_{i_{*}}$ since we can safely assume that marginal utilities of additional ratings or slightly higher means could diminish.\par

Matrix factorization methods are used in recommender systems to derive a set of latent factors, from the user $\times$ item rating matrix, to characterize both users and items by this vector of latent factors. The user-item interaction is modeled as the dot product of the latent factor space \citep{Koren2009}. Accordingly, in the base version of a rating prediction algorithm each item \emph{j} will be associated with a vector of factors $q_j$, and each user $i$ is associated with a vector of factors $p_i$ and predictions can be derived from the dot product of their factor vectors $\hat{r}_{ui} =p_i \ast q_j^T$.
Thus, in the equation below we add our utility weights $u_{ij}$ as a soft constraint in analogy to \cite{Abdollahi2017UsingFactorization}: 
 
\begin{equation}
\label{eq:novelMF}
\sum_{i, j \in R}(r_{ij} - {p_{i}q^T_{j}})^2 + 
\frac{\phi}{2} {(\|p_{i}\|^2 + \|q_{j}\|^2)} + \frac{\delta}{2}||p_i-q_j||^2u_{ij}
\end{equation}
where $\phi$ and $\delta$ are regularization coefficients, and $u_{ij}$ is the user's $i$ perceived utility of item $j$'s rating summarization.
We use a $L_2$ regularization term to properly fit the model to the data. To minimize the observed loss function of Formula~\ref{eq:novelMF}, we used a stochastic gradient descent.

\begin{figure}
\centering
    \begin{subfigure}[b]{0.5\textwidth}
        \centering
		\includegraphics[scale=0.40]{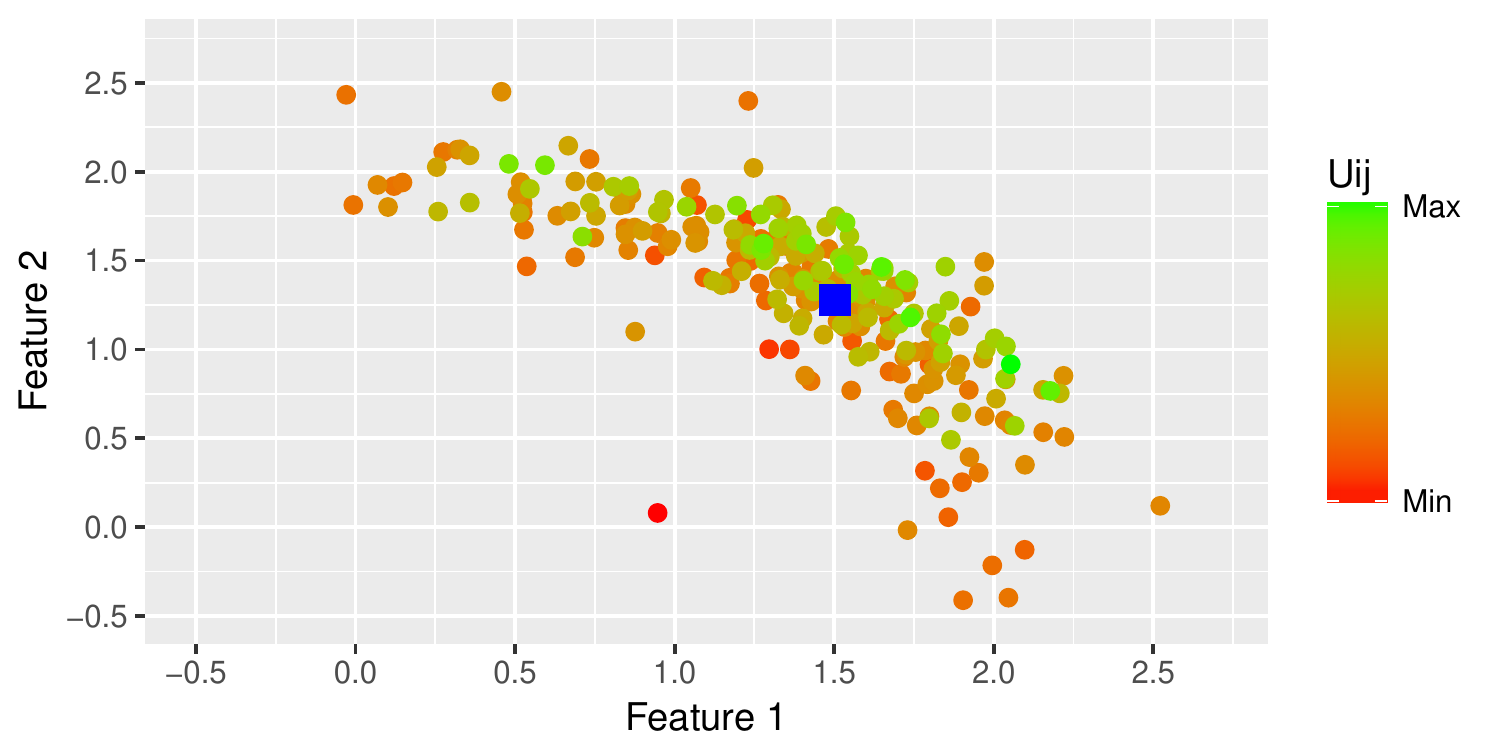}
        \caption{Matrix Factorization}
        \label{img:mf}
    \end{subfigure}%
    
    \begin{subfigure}[b]{0.5\textwidth}
        \centering
		\includegraphics[scale=0.40]{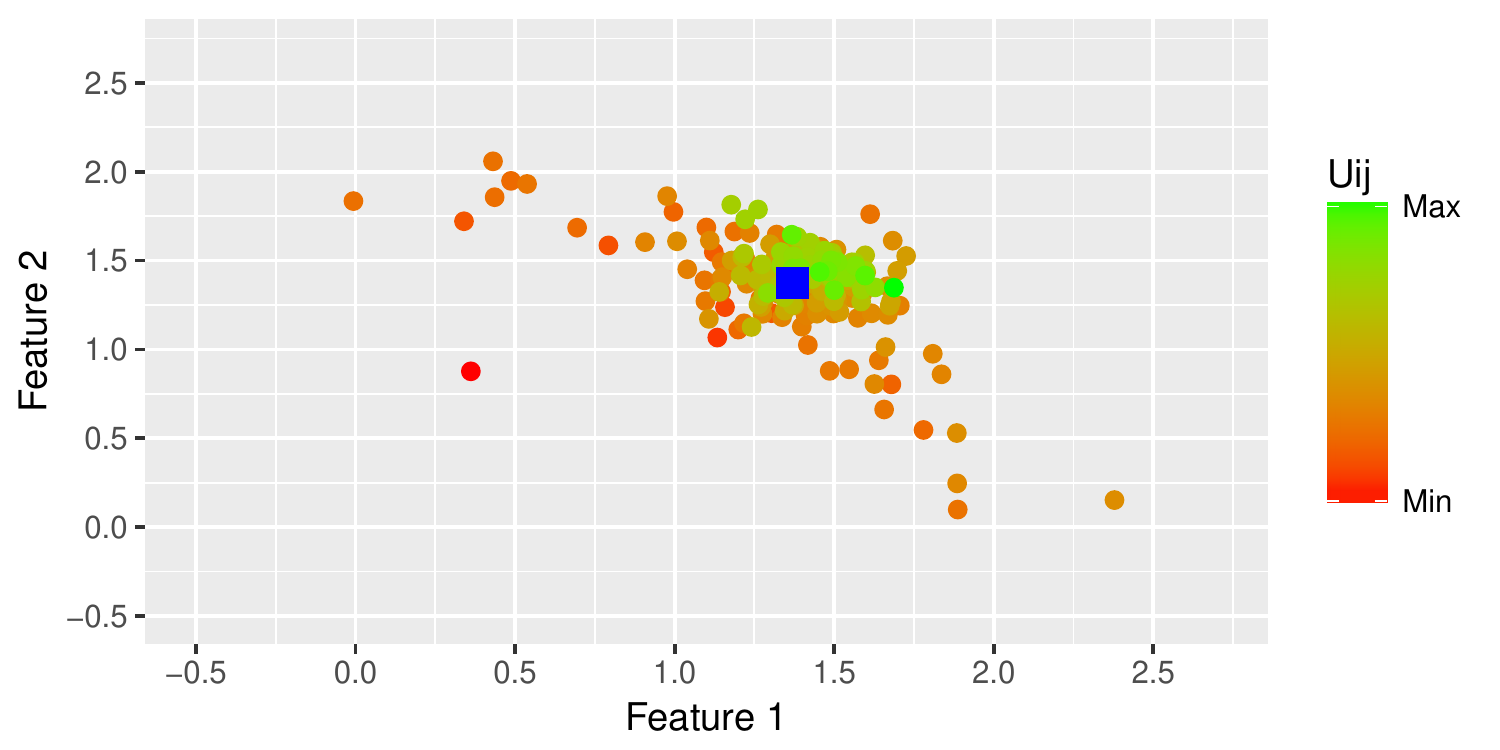}
        \caption{Utility-aware Matrix Factorization}
        \label{img:weighted_mf}
    \end{subfigure}
    \caption{Matrix factorization in the two-dimensional latent space. For a given random user from the TripAdvisor dataset, represented by the blue square, green dots are a graphical representation of items with the highest measured utility in the two dimensional latent space. The red dots are the lowest measured utility for a given user.}
\label{fig:mf_ls}
\end{figure}

In a toy example, we trained both the base MF and the constrained MF on our tourism dataset, by setting the latent space equal to two dimensions, and project users and items onto the two-dimensional latent space, as anecdotal evidence for the functioning of the approach.
The blue square depicts a randomly drawn user from the dataset. Green dots are explanations that users are more likely to accept, according to the utility function in Formula~\ref{eq:formula}. In Figure~\ref{img:mf} the green dots are spread all over the latent space. After applying the soft constraint to the MF, Figure~\ref{img:weighted_mf}, all explainable items are shifted closer to the user. Obviously, using more features would lead to better results as well as application of this idea to different (ranking) algorithms remains to be demonstrated. For this example, however, we used a prediction algorithm  for the sake of easier visualization and due to the recent work of \cite{Abdollahi2017UsingFactorization}.

\subsection{Considerations on personality results}
Scholars in behavioral research developed considerable evidence for the existence of individual differences in the desire to maximize or satisfice \cite{dar2009maximization,iyengar2006doing,misuraca2013time,schwartz2002maximizing,sparks2012failing}, but, so far, attempts at applying these insights onto the recommender systems domain yielded inconclusive results \cite{Jugovac2018InvestigatingRecommendations,Knijnenburg2011EachSystems}. In the present study, we took on the suggestion from \cite{cheek2016meaning,Nenkov2008AStudies} that maximization is better understood by looking into overall maximization as well as into its sub-components. This enabled us to confirm the body of evidence from behavioral studies stating that maximizers respond differently to rating summarizations than satisficers. Moreover, we observed that this behavioral difference among maximizers is due to their difficulty in making their choices when exposed to various rating summarizations. Apparently, they suffer more heavily from what is known as choice overload \cite{scheibehenne2010can}, and this insecurity to discover "the best possible choice" leads maximizers with decision difficulty to respond to rating summarizations in a highly distinctive manner. These findings are valuable to the domain of recommender systems, since they can lead to the development of adaptation and personalization strategies that would help to lower the perceived decision difficulty, or even the potential experience of regret.

\section{Conclusions }

This paper presented a choice-based conjoint (CBC) experiment that explored how different summarizations of rating distributions, like the total number of ratings, mean, variance, skewness or the origin of the ratings themselves, impact users' decision making. By putting attribute levels that are not only representative for the tourism domain, but also for e-commerce in general, to test, we noticed that users are willing to trade-in an alternative with a higher mean rating value based only on few ratings against an item with a lower mean that is based on many more ratings. Importantly, this behavior is moderated by the decision style of participants, where people with a high behavioral tendency towards maximization, and in particular those high on the decision difficulty dimension underlying maximizing tendencies, still rely primarily on high mean values. In contrast, their counterparts that do not experience decision difficulty are more free to weight in different characteristics of rating distributions against each other. These results require a more fine-grained sensitivity analysis as future work in order to serve as a basis for tuning recommendation algorithms according to users' presumed decision making styles.  

\bibliographystyle{ACM-Reference-Format}
\bibliography{bibliography,Mendeley}


\begin{thebibliography}{00}


\ifx \showCODEN    \undefined \def \showCODEN     #1{\unskip}     \fi
\ifx \showDOI      \undefined \def \showDOI       #1{#1}\fi
\ifx \showISBNx    \undefined \def \showISBNx     #1{\unskip}     \fi
\ifx \showISBNxiii \undefined \def \showISBNxiii  #1{\unskip}     \fi
\ifx \showISSN     \undefined \def \showISSN      #1{\unskip}     \fi
\ifx \showLCCN     \undefined \def \showLCCN      #1{\unskip}     \fi
\ifx \shownote     \undefined \def \shownote      #1{#1}          \fi
\ifx \showarticletitle \undefined \def \showarticletitle #1{#1}   \fi
\ifx \showURL      \undefined \def \showURL       {\relax}        \fi
\providecommand\bibfield[2]{#2}
\providecommand\bibinfo[2]{#2}
\providecommand\natexlab[1]{#1}
\providecommand\showeprint[2][]{arXiv:#2}

\bibitem[\protect\citeauthoryear{Abdollahi and Nasraoui}{Abdollahi and
  Nasraoui}{2017}]%
        {Abdollahi2017UsingFactorization}
\bibfield{author}{\bibinfo{person}{Behnoush Abdollahi} {and}
  \bibinfo{person}{Olfa Nasraoui}.} \bibinfo{year}{2017}\natexlab{}.
\newblock \showarticletitle{{Using Explainability for Constrained Matrix
  Factorization}}. In \bibinfo{booktitle}{{\em Proceedings of the Eleventh ACM
  Conference on Recommender Systems - RecSys '17}}. \bibinfo{pages}{79--83}.
\newblock
\showISBNx{9781450346528}
\showDOI{%
\url{https://doi.org/10.1145/3109859.3109913}}


\bibitem[\protect\citeauthoryear{Agarwal, DeSarbo, Malhotra, and Rao}{Agarwal
  et~al\mbox{.}}{2015}]%
        {agarwal2015interdisciplinary}
\bibfield{author}{\bibinfo{person}{James Agarwal}, \bibinfo{person}{Wayne~S
  DeSarbo}, \bibinfo{person}{Naresh~K Malhotra}, {and}
  \bibinfo{person}{Vithala~R Rao}.} \bibinfo{year}{2015}\natexlab{}.
\newblock \showarticletitle{An interdisciplinary review of research in conjoint
  analysis: recent developments and directions for future research}.
\newblock \bibinfo{journal}{{\em Customer Needs and Solutions\/}}
  \bibinfo{volume}{2}, \bibinfo{number}{1} (\bibinfo{year}{2015}),
  \bibinfo{pages}{19--40}.
\newblock


\bibitem[\protect\citeauthoryear{Bilgic and Mooney}{Bilgic and Mooney}{2005}]%
        {Bilgic2005ExplainingPromotion}
\bibfield{author}{\bibinfo{person}{Mustafa Bilgic} {and}
  \bibinfo{person}{Raymond~J Mooney}.} \bibinfo{year}{2005}\natexlab{}.
\newblock \showarticletitle{{Explaining Recommendations: Satisfaction vs.
  Promotion}}.
\newblock \bibinfo{journal}{{\em Proceedings of Beyond Personalization 2005: A
  Workshop on the Next Stage of Recommender Systems Research at The 2005
  International Conference on Intelligent User Interfaces\/}}
  (\bibinfo{year}{2005}), \bibinfo{pages}{13--18}.
\newblock
\showISBNx{1581138946}
\showDOI{%
\url{https://doi.org/10.1145/1040830.1040839}}


\bibitem[\protect\citeauthoryear{Carbonell and Brand}{Carbonell and
  Brand}{2018}]%
        {Carbonell2018ChoosingPhysician}
\bibfield{author}{\bibinfo{person}{Guillermo Carbonell} {and}
  \bibinfo{person}{Matthias Brand}.} \bibinfo{year}{2018}\natexlab{}.
\newblock \showarticletitle{{Choosing a Physician on Social Media: Comments and
  Ratings of Users are More Important than the Qualification of a Physician}}.
\newblock \bibinfo{journal}{{\em International Journal of Human-Computer
  Interaction\/}} \bibinfo{volume}{34}, \bibinfo{number}{2} (\bibinfo{date}{2}
  \bibinfo{year}{2018}), \bibinfo{pages}{117--128}.
\newblock
\showISSN{15327590}
\showDOI{%
\url{https://doi.org/10.1080/10447318.2017.1330803}}


\bibitem[\protect\citeauthoryear{Cheek and Schwartz}{Cheek and
  Schwartz}{2016}]%
        {cheek2016meaning}
\bibfield{author}{\bibinfo{person}{Nathan~N Cheek} {and} \bibinfo{person}{Barry
  Schwartz}.} \bibinfo{year}{2016}\natexlab{}.
\newblock \showarticletitle{On the meaning and measurement of maximization}.
\newblock \bibinfo{journal}{{\em Judgment and Decision making\/}}
  \bibinfo{volume}{11}, \bibinfo{number}{2} (\bibinfo{year}{2016}),
  \bibinfo{pages}{126}.
\newblock


\bibitem[\protect\citeauthoryear{Chen, de~Gemmis, Felfernig, Lops, Ricci, and
  Semeraro}{Chen et~al\mbox{.}}{2013}]%
        {chen2013human}
\bibfield{author}{\bibinfo{person}{Li Chen}, \bibinfo{person}{Marco de Gemmis},
  \bibinfo{person}{Alexander Felfernig}, \bibinfo{person}{Pasquale Lops},
  \bibinfo{person}{Francesco Ricci}, {and} \bibinfo{person}{Giovanni
  Semeraro}.} \bibinfo{year}{2013}\natexlab{}.
\newblock \showarticletitle{Human decision making and recommender systems}.
\newblock \bibinfo{journal}{{\em ACM Transactions on Interactive Intelligent
  Systems (TiiS)\/}} \bibinfo{volume}{3}, \bibinfo{number}{3}
  (\bibinfo{year}{2013}), \bibinfo{pages}{17}.
\newblock


\bibitem[\protect\citeauthoryear{Cho, Kwon, Na, Suk, and Lee}{Cho
  et~al\mbox{.}}{2015}]%
        {Cho2015TheStyle}
\bibfield{author}{\bibinfo{person}{Minji Cho}, \bibinfo{person}{Soyoung Kwon},
  \bibinfo{person}{Nooree Na}, \bibinfo{person}{Hyeon-Jeong Suk}, {and}
  \bibinfo{person}{Kun-Pyo Lee}.} \bibinfo{year}{2015}\natexlab{}.
\newblock \showarticletitle{{The Elders Preference for Skeuomorphism as App
  Icon Style}}.
\newblock \bibinfo{journal}{{\em Proceedings of the 33rd Annual ACM Conference
  Extended Abstracts on Human Factors in Computing Systems - CHI EA '15\/}}
  (\bibinfo{year}{2015}), \bibinfo{pages}{899--904}.
\newblock
\showISBNx{9781450331463}
\showDOI{%
\url{https://doi.org/10.1145/2702613.2732887}}


\bibitem[\protect\citeauthoryear{Chung and Rao}{Chung and Rao}{2012}]%
        {chung2012general}
\bibfield{author}{\bibinfo{person}{Jaihak Chung} {and}
  \bibinfo{person}{Vithala~R Rao}.} \bibinfo{year}{2012}\natexlab{}.
\newblock \showarticletitle{A general consumer preference model for experience
  products: application to internet recommendation services}.
\newblock \bibinfo{journal}{{\em Journal of marketing research\/}}
  \bibinfo{volume}{49}, \bibinfo{number}{3} (\bibinfo{year}{2012}),
  \bibinfo{pages}{289--305}.
\newblock


\bibitem[\protect\citeauthoryear{Cosley, Lam, Albert, Konstan, and
  Riedl}{Cosley et~al\mbox{.}}{2003}]%
        {Cosley2003}
\bibfield{author}{\bibinfo{person}{Dan Cosley}, \bibinfo{person}{Shyong~K.
  Lam}, \bibinfo{person}{Istvan Albert}, \bibinfo{person}{Joseph~a. Konstan},
  {and} \bibinfo{person}{John Riedl}.} \bibinfo{year}{2003}\natexlab{}.
\newblock \showarticletitle{{Is Seeing Believing? How Recommender System
  Interfaces Affect Users' Opinions}}.
\newblock \bibinfo{journal}{{\em Proceedings of the Conference on Human Factors
  in Computing Systems (CHI '03)\/}} \bibinfo{number}{5}
  (\bibinfo{year}{2003}), \bibinfo{pages}{585--592}.
\newblock
\showISBNx{1581136307}
\showDOI{%
\url{https://doi.org/10.1145/642611.642713}}


\bibitem[\protect\citeauthoryear{Dar-Nimrod, Rawn, Lehman, and
  Schwartz}{Dar-Nimrod et~al\mbox{.}}{2009}]%
        {dar2009maximization}
\bibfield{author}{\bibinfo{person}{Ilan Dar-Nimrod},
  \bibinfo{person}{Catherine~D Rawn}, \bibinfo{person}{Darrin~R Lehman}, {and}
  \bibinfo{person}{Barry Schwartz}.} \bibinfo{year}{2009}\natexlab{}.
\newblock \showarticletitle{The maximization paradox: The costs of seeking
  alternatives}.
\newblock \bibinfo{journal}{{\em Personality and Individual Differences\/}}
  \bibinfo{volume}{46}, \bibinfo{number}{5-6} (\bibinfo{year}{2009}),
  \bibinfo{pages}{631--635}.
\newblock


\bibitem[\protect\citeauthoryear{de~Langhe, Fernbach, and
  Lichtenstein}{de~Langhe et~al\mbox{.}}{2016}]%
        {deLanghe2016NavigatingRatings}
\bibfield{author}{\bibinfo{person}{Bart de Langhe}, \bibinfo{person}{Philip~M.
  Fernbach}, {and} \bibinfo{person}{Donald~R. Lichtenstein}.}
  \bibinfo{year}{2016}\natexlab{}.
\newblock \showarticletitle{{Navigating by the stars: Investigating the actual
  and perceived validity of online user ratings}}.
\newblock \bibinfo{journal}{{\em Journal of Consumer Research\/}}
  \bibinfo{volume}{42}, \bibinfo{number}{6} (\bibinfo{year}{2016}),
  \bibinfo{pages}{817--833}.
\newblock
\showISSN{00935301}
\showDOI{%
\url{https://doi.org/10.1093/jcr/ucv047}}


\bibitem[\protect\citeauthoryear{Duan, Gu, and Whinston}{Duan
  et~al\mbox{.}}{2008}]%
        {duan2008online}
\bibfield{author}{\bibinfo{person}{Wenjing Duan}, \bibinfo{person}{Bin Gu},
  {and} \bibinfo{person}{Andrew~B Whinston}.} \bibinfo{year}{2008}\natexlab{}.
\newblock \showarticletitle{Do online reviews matter?-An empirical
  investigation of panel data}.
\newblock \bibinfo{journal}{{\em Decision support systems\/}}
  \bibinfo{volume}{45}, \bibinfo{number}{4} (\bibinfo{year}{2008}),
  \bibinfo{pages}{1007--1016}.
\newblock


\bibitem[\protect\citeauthoryear{Friedrich and Zanker}{Friedrich and
  Zanker}{2011}]%
        {Friedrich2011ASystems}
\bibfield{author}{\bibinfo{person}{Gerhard Friedrich} {and}
  \bibinfo{person}{Markus Zanker}.} \bibinfo{year}{2011}\natexlab{}.
\newblock \showarticletitle{{A Taxonomy for Generating Explanations in
  Recommender Systems}}.
\newblock \bibinfo{journal}{{\em AI Magazine\/}} \bibinfo{volume}{32},
  \bibinfo{number}{3} (\bibinfo{year}{2011}), \bibinfo{pages}{90}.
\newblock
\showISSN{0738-4602}
\showDOI{%
\url{https://doi.org/10.1609/aimag.v32i3.2365}}


\bibitem[\protect\citeauthoryear{Fuchs and Zanker}{Fuchs and Zanker}{2012}]%
        {Fuchs2012Multi-criteriaDomain}
\bibfield{author}{\bibinfo{person}{Matthias Fuchs} {and}
  \bibinfo{person}{Markus Zanker}.} \bibinfo{year}{2012}\natexlab{}.
\newblock \showarticletitle{{Multi-criteria ratings for recommender systems: An
  empirical analysis in the tourism domain}}. In \bibinfo{booktitle}{{\em
  Lecture Notes in Business Information Processing}}, Vol.~\bibinfo{volume}{123
  LNBIP}. \bibinfo{pages}{100--111}.
\newblock
\showISBNx{9783642322723}
\showISSN{18651348}
\showDOI{%
\url{https://doi.org/10.1007/978-3-642-32273-0{\_}9}}


\bibitem[\protect\citeauthoryear{Green and Srinivasan}{Green and
  Srinivasan}{1990}]%
        {green1990conjoint}
\bibfield{author}{\bibinfo{person}{Paul~E Green} {and} \bibinfo{person}{Venkat
  Srinivasan}.} \bibinfo{year}{1990}\natexlab{}.
\newblock \showarticletitle{Conjoint analysis in marketing: new developments
  with implications for research and practice}.
\newblock \bibinfo{journal}{{\em The journal of marketing\/}}
  (\bibinfo{year}{1990}), \bibinfo{pages}{3--19}.
\newblock


\bibitem[\protect\citeauthoryear{Hauber, Gonz{\'{a}}lez, Groothuis-oudshoorn,
  Prior, Marshall, Cunningham, Ijzerman, and Bridges}{Hauber
  et~al\mbox{.}}{2016}]%
        {Hauber2016StatisticalForce}
\bibfield{author}{\bibinfo{person}{A.~Brett Hauber},
  \bibinfo{person}{Juan~Marcos Gonz{\'{a}}lez}, \bibinfo{person}{Catharina
  G.M.~M Groothuis-oudshoorn}, \bibinfo{person}{Thomas Prior},
  \bibinfo{person}{Deborah~A. Marshall}, \bibinfo{person}{Charles Cunningham},
  \bibinfo{person}{Maarten~J. Ijzerman}, {and} \bibinfo{person}{John F.P.~P
  Bridges}.} \bibinfo{year}{2016}\natexlab{}.
\newblock \showarticletitle{{Statistical Methods for the Analysis of Discrete
  Choice Experiments: A Report of the ISPOR Conjoint Analysis Good Research
  Practices Task Force}}.
\newblock \bibinfo{journal}{{\em Value in Health\/}} \bibinfo{volume}{19},
  \bibinfo{number}{4} (\bibinfo{year}{2016}), \bibinfo{pages}{300--315}.
\newblock
\showISBNx{1098-3015}
\showISSN{15244733}
\showDOI{%
\url{https://doi.org/10.1016/j.jval.2016.04.004}}


\bibitem[\protect\citeauthoryear{Hauser and Rao}{Hauser and Rao}{2004}]%
        {hauser2004conjoint}
\bibfield{author}{\bibinfo{person}{John~R Hauser} {and}
  \bibinfo{person}{Vithala~R Rao}.} \bibinfo{year}{2004}\natexlab{}.
\newblock \showarticletitle{Conjoint analysis, related modeling, and
  applications}.
\newblock In \bibinfo{booktitle}{{\em Marketing Research and Modeling: Progress
  and Prospects}}. \bibinfo{publisher}{Springer}, \bibinfo{pages}{141--168}.
\newblock


\bibitem[\protect\citeauthoryear{Herlocker, Konstan, and Riedl}{Herlocker
  et~al\mbox{.}}{2000}]%
        {Herlocker2000ExplainingRecommendations}
\bibfield{author}{\bibinfo{person}{Jonathan~L Herlocker},
  \bibinfo{person}{Joseph~A Konstan}, {and} \bibinfo{person}{John Riedl}.}
  \bibinfo{year}{2000}\natexlab{}.
\newblock \showarticletitle{{Explaining collaborative filtering
  recommendations}}. In \bibinfo{booktitle}{{\em Proceedings of the 2000 ACM
  conference on Computer supported cooperative work - CSCW '00}}.
  \bibinfo{pages}{241--250}.
\newblock
\showISBNx{1581132220}
\showISSN{00318655}
\showDOI{%
\url{https://doi.org/10.1145/358916.358995}}


\bibitem[\protect\citeauthoryear{Iyengar, Wells, and Schwartz}{Iyengar
  et~al\mbox{.}}{2006}]%
        {iyengar2006doing}
\bibfield{author}{\bibinfo{person}{Sheena~S Iyengar},
  \bibinfo{person}{Rachael~E Wells}, {and} \bibinfo{person}{Barry Schwartz}.}
  \bibinfo{year}{2006}\natexlab{}.
\newblock \showarticletitle{Doing better but feeling worse: Looking for the
  “best” job undermines satisfaction}.
\newblock \bibinfo{journal}{{\em Psychological Science\/}}
  \bibinfo{volume}{17}, \bibinfo{number}{2} (\bibinfo{year}{2006}),
  \bibinfo{pages}{143--150}.
\newblock


\bibitem[\protect\citeauthoryear{Jannach, Lerche, and Jugovac}{Jannach
  et~al\mbox{.}}{2015}]%
        {Jannach2015}
\bibfield{author}{\bibinfo{person}{Dietmar Jannach}, \bibinfo{person}{Lukas
  Lerche}, {and} \bibinfo{person}{Michael Jugovac}.}
  \bibinfo{year}{2015}\natexlab{}.
\newblock \showarticletitle{{Item familiarity as a possible confounding factor
  in user-centric recommender systems evaluation}}.
\newblock \bibinfo{journal}{{\em I-Com\/}} \bibinfo{volume}{14},
  \bibinfo{number}{1} (\bibinfo{year}{2015}), \bibinfo{pages}{29--39}.
\newblock
\showURL{%
\url{http://dx.doi.org/10.1515/icom-2015-0018}}


\bibitem[\protect\citeauthoryear{Johnson, Lancsar, Marshall, Kilambi,
  M{\"{u}}hlbacher, Regier, Bresnahan, Kanninen, and Bridges}{Johnson
  et~al\mbox{.}}{2013}]%
        {Johnson2013ConstructingForce}
\bibfield{author}{\bibinfo{person}{F.~Reed Johnson}, \bibinfo{person}{Emily
  Lancsar}, \bibinfo{person}{Deborah Marshall}, \bibinfo{person}{Vikram
  Kilambi}, \bibinfo{person}{Axel M{\"{u}}hlbacher}, \bibinfo{person}{Dean~A.
  Regier}, \bibinfo{person}{Brian~W. Bresnahan}, \bibinfo{person}{Barbara
  Kanninen}, {and} \bibinfo{person}{John~F.P. Bridges}.}
  \bibinfo{year}{2013}\natexlab{}.
\newblock \showarticletitle{{Constructing experimental designs for
  discrete-choice experiments: Report of the ISPOR conjoint analysis
  experimental design good research practices task force}}.
\newblock \bibinfo{journal}{{\em Value in Health\/}} \bibinfo{volume}{16},
  \bibinfo{number}{1} (\bibinfo{date}{1} \bibinfo{year}{2013}),
  \bibinfo{pages}{3--13}.
\newblock
\showISBNx{1098-3015}
\showISSN{10983015}
\showDOI{%
\url{https://doi.org/10.1016/j.jval.2012.08.2223}}


\bibitem[\protect\citeauthoryear{Jugovac, Nunes, and Jannach}{Jugovac
  et~al\mbox{.}}{2018}]%
        {Jugovac2018InvestigatingRecommendations}
\bibfield{author}{\bibinfo{person}{Michael Jugovac}, \bibinfo{person}{Ingrid
  Nunes}, {and} \bibinfo{person}{Dietmar Jannach}.}
  \bibinfo{year}{2018}\natexlab{}.
\newblock \showarticletitle{{Investigating the Decision-Making Behavior of
  Maximizers and Satisficers in the Presence of Recommendations}}. In
  \bibinfo{booktitle}{{\em 26th Conference on User Modeling, Adaptation and
  Personalization (UMAP'18)}}.
\newblock
\showISBNx{1234567245}
\showDOI{%
\url{https://doi.org/10.475/123}}


\bibitem[\protect\citeauthoryear{Knijnenburg and Willemsen}{Knijnenburg and
  Willemsen}{2011}]%
        {Knijnenburg2011EachSystems}
\bibfield{author}{\bibinfo{person}{Bart~P Knijnenburg} {and}
  \bibinfo{person}{Martijn~C Willemsen}.} \bibinfo{year}{2011}\natexlab{}.
\newblock \showarticletitle{{Each to His Own : How Different Users Call for
  Different Interaction Methods in Recommender Systems}}.
\newblock \bibinfo{journal}{{\em Proceedings of the 5th ACM conference on
  Recommender systems - RecSys '11\/}} (\bibinfo{year}{2011}),
  \bibinfo{pages}{141--148}.
\newblock
\showISBNx{9781450306836}
\showDOI{%
\url{https://doi.org/10.1145/2043932.2043960}}


\bibitem[\protect\citeauthoryear{Koren, Bell, and Volinsky}{Koren
  et~al\mbox{.}}{2009}]%
        {Koren2009}
\bibfield{author}{\bibinfo{person}{Y. Koren}, \bibinfo{person}{R. Bell}, {and}
  \bibinfo{person}{C. Volinsky}.} \bibinfo{year}{2009}\natexlab{}.
\newblock \showarticletitle{{Matrix Factorization Techniques for Recommender
  Systems}}.
\newblock \bibinfo{journal}{{\em Computer\/}} \bibinfo{volume}{42},
  \bibinfo{number}{8} (\bibinfo{year}{2009}), \bibinfo{pages}{42--49}.
\newblock
\showISBNx{0018-9162}
\showISSN{0018-9162}
\showDOI{%
\url{https://doi.org/10.1109/MC.2009.263}}


\bibitem[\protect\citeauthoryear{Kouki, Schaffer, Pujara, O'Donovan, and
  Getoor}{Kouki et~al\mbox{.}}{2017}]%
        {Kouki2017UserExplanations}
\bibfield{author}{\bibinfo{person}{Pigi Kouki}, \bibinfo{person}{James
  Schaffer}, \bibinfo{person}{Jay Pujara}, \bibinfo{person}{John O'Donovan},
  {and} \bibinfo{person}{Lise Getoor}.} \bibinfo{year}{2017}\natexlab{}.
\newblock \showarticletitle{{User Preferences for Hybrid Explanations}}. In
  \bibinfo{booktitle}{{\em Proceedings of the Eleventh ACM Conference on
  Recommender Systems - RecSys '17}}. \bibinfo{pages}{84--88}.
\newblock
\showISBNx{9781450346528}
\showDOI{%
\url{https://doi.org/10.1145/3109859.3109915}}


\bibitem[\protect\citeauthoryear{Kuhfeld}{Kuhfeld}{2005}]%
        {Kuhfeld2010}
\bibfield{author}{\bibinfo{person}{Warren Kuhfeld}.}
  \bibinfo{year}{2005}\natexlab{}.
\newblock \showarticletitle{{Experimental design, efficiency, coding, and
  choice designs}}.
\newblock \bibinfo{journal}{{\em Marketing research methods in sas:
  Experimental design, choice, conjoint, and graphical techniques\/}}
  (\bibinfo{year}{2005}), \bibinfo{pages}{47--97}.
\newblock
\showURL{%
\url{https://support.sas.com/techsup/technote/mr2010c.pdf}}


\bibitem[\protect\citeauthoryear{Kuhfeld}{Kuhfeld}{2010}]%
        {Kuhfeld2010DiscreteChoice}
\bibfield{author}{\bibinfo{person}{Warren~F Kuhfeld}.}
  \bibinfo{year}{2010}\natexlab{}.
\newblock \showarticletitle{{Discrete Choice}}.
\newblock \bibinfo{journal}{{\em SAS Technical Papers\/}}
  \bibinfo{volume}{MR-2010F} (\bibinfo{year}{2010}), \bibinfo{pages}{285--663}.
\newblock
\showISBNx{0131395386}
\showURL{%
\url{http://support.sas.com/techsup/technote/mr2010f.pdf}}


\bibitem[\protect\citeauthoryear{Louviere, Flynn, and Carson}{Louviere
  et~al\mbox{.}}{2010}]%
        {louviere2010discrete}
\bibfield{author}{\bibinfo{person}{Jordan~J Louviere}, \bibinfo{person}{Terry~N
  Flynn}, {and} \bibinfo{person}{Richard~T Carson}.}
  \bibinfo{year}{2010}\natexlab{}.
\newblock \showarticletitle{Discrete choice experiments are not conjoint
  analysis}.
\newblock \bibinfo{journal}{{\em Journal of Choice Modelling\/}}
  \bibinfo{volume}{3}, \bibinfo{number}{3} (\bibinfo{year}{2010}),
  \bibinfo{pages}{57--72}.
\newblock


\bibitem[\protect\citeauthoryear{Misuraca and Teuscher}{Misuraca and
  Teuscher}{2013}]%
        {misuraca2013time}
\bibfield{author}{\bibinfo{person}{Raffaella Misuraca} {and}
  \bibinfo{person}{Ursina Teuscher}.} \bibinfo{year}{2013}\natexlab{}.
\newblock \showarticletitle{Time flies when you maximize—Maximizers and
  satisficers perceive time differently when making decisions}.
\newblock \bibinfo{journal}{{\em Acta psychologica\/}} \bibinfo{volume}{143},
  \bibinfo{number}{2} (\bibinfo{year}{2013}), \bibinfo{pages}{176--180}.
\newblock


\bibitem[\protect\citeauthoryear{Nenkov, Morrin, Ward, Schwartz, and
  Hulland}{Nenkov et~al\mbox{.}}{2008a}]%
        {Nenkov2008AStudies}
\bibfield{author}{\bibinfo{person}{Gergana~Y Nenkov}, \bibinfo{person}{Maureen
  Morrin}, \bibinfo{person}{Andrew Ward}, \bibinfo{person}{Barry Schwartz},
  {and} \bibinfo{person}{John Hulland}.} \bibinfo{year}{2008}\natexlab{a}.
\newblock \showarticletitle{{A short form of the Maximization Scale: Factor
  structure, reliability and validity studies}}.
\newblock \bibinfo{journal}{{\em Judgment and Decision Making\/}}
  \bibinfo{volume}{3}, \bibinfo{number}{5} (\bibinfo{year}{2008}),
  \bibinfo{pages}{371--388}.
\newblock
\showURL{%
\url{http://www.sjdm.org/journal/8323/jdm8323.pdf}}


\bibitem[\protect\citeauthoryear{Nenkov, Morrin, Ward, Schwartz, and
  Hulland}{Nenkov et~al\mbox{.}}{2008b}]%
        {nenkov2008short}
\bibfield{author}{\bibinfo{person}{Gergana~Y Nenkov}, \bibinfo{person}{Maureen
  Morrin}, \bibinfo{person}{Andrew Ward}, \bibinfo{person}{Barry Schwartz},
  {and} \bibinfo{person}{John Hulland}.} \bibinfo{year}{2008}\natexlab{b}.
\newblock \showarticletitle{A short form of the Maximization Scale: Factor
  structure, reliability and validity studies}.
\newblock \bibinfo{journal}{{\em Judgment and Decision Making\/}}
  \bibinfo{volume}{3}, \bibinfo{number}{5} (\bibinfo{year}{2008}),
  \bibinfo{pages}{371--388}.
\newblock


\bibitem[\protect\citeauthoryear{Nunes and Jannach}{Nunes and Jannach}{2017}]%
        {Nunes2017ASystems}
\bibfield{author}{\bibinfo{person}{Ingrid Nunes} {and} \bibinfo{person}{Dietmar
  Jannach}.} \bibinfo{year}{2017}\natexlab{}.
\newblock \showarticletitle{{A systematic review and taxonomy of explanations
  in decision support and recommender systems}}.
\newblock \bibinfo{journal}{{\em User Modeling and User-Adapted Interaction\/}}
  \bibinfo{volume}{27}, \bibinfo{number}{3-5} (\bibinfo{date}{12}
  \bibinfo{year}{2017}), \bibinfo{pages}{393--444}.
\newblock
\showISSN{15731391}
\showDOI{%
\url{https://doi.org/10.1007/s11257-017-9195-0}}


\bibitem[\protect\citeauthoryear{Rao}{Rao}{2008}]%
        {rao2008developments}
\bibfield{author}{\bibinfo{person}{Vithala~R Rao}.}
  \bibinfo{year}{2008}\natexlab{}.
\newblock \showarticletitle{Developments in conjoint analysis}.
\newblock In \bibinfo{booktitle}{{\em Handbook of marketing decision models}}.
  \bibinfo{publisher}{Springer}, \bibinfo{pages}{23--53}.
\newblock


\bibitem[\protect\citeauthoryear{Rao}{Rao}{2014}]%
        {Rao2014ChoiceAnalysis}
\bibfield{author}{\bibinfo{person}{Vithala~R Rao}.}
  \bibinfo{year}{2014}\natexlab{}.
\newblock \showarticletitle{{Choice Based Conjoint Studies: Design and
  Analysis}}.
\newblock In \bibinfo{booktitle}{{\em Applied Conjoint Analysis}}.
  \bibinfo{pages}{127--183}.
\newblock
\showISBNx{978-3-540-87752-3 978-3-540-87753-0}
\showDOI{%
\url{https://doi.org/10.1007/978-3-540-87753-0{\_}4}}


\bibitem[\protect\citeauthoryear{Scheibehenne, Greifeneder, and
  Todd}{Scheibehenne et~al\mbox{.}}{2010}]%
        {scheibehenne2010can}
\bibfield{author}{\bibinfo{person}{Benjamin Scheibehenne},
  \bibinfo{person}{Rainer Greifeneder}, {and} \bibinfo{person}{Peter~M Todd}.}
  \bibinfo{year}{2010}\natexlab{}.
\newblock \showarticletitle{Can there ever be too many options? A meta-analytic
  review of choice overload}.
\newblock \bibinfo{journal}{{\em Journal of Consumer Research\/}}
  \bibinfo{volume}{37}, \bibinfo{number}{3} (\bibinfo{year}{2010}),
  \bibinfo{pages}{409--425}.
\newblock


\bibitem[\protect\citeauthoryear{Schwartz, Ward, Monterosso, Lyubomirsky,
  White, and Lehman}{Schwartz et~al\mbox{.}}{2002}]%
        {schwartz2002maximizing}
\bibfield{author}{\bibinfo{person}{Barry Schwartz}, \bibinfo{person}{Andrew
  Ward}, \bibinfo{person}{John Monterosso}, \bibinfo{person}{Sonja
  Lyubomirsky}, \bibinfo{person}{Katherine White}, {and}
  \bibinfo{person}{Darrin~R Lehman}.} \bibinfo{year}{2002}\natexlab{}.
\newblock \showarticletitle{Maximizing versus satisficing: Happiness is a
  matter of choice.}
\newblock \bibinfo{journal}{{\em Journal of personality and social
  psychology\/}} \bibinfo{volume}{83}, \bibinfo{number}{5}
  (\bibinfo{year}{2002}), \bibinfo{pages}{1178}.
\newblock


\bibitem[\protect\citeauthoryear{Simon}{Simon}{1955}]%
        {simon1955behavioral}
\bibfield{author}{\bibinfo{person}{Herbert~A Simon}.}
  \bibinfo{year}{1955}\natexlab{}.
\newblock \showarticletitle{A behavioral model of rational choice}.
\newblock \bibinfo{journal}{{\em The quarterly journal of economics\/}}
  \bibinfo{volume}{69}, \bibinfo{number}{1} (\bibinfo{year}{1955}),
  \bibinfo{pages}{99--118}.
\newblock


\bibitem[\protect\citeauthoryear{Sparks, Ehrlinger, and Eibach}{Sparks
  et~al\mbox{.}}{2012}]%
        {sparks2012failing}
\bibfield{author}{\bibinfo{person}{Erin~A Sparks}, \bibinfo{person}{Joyce
  Ehrlinger}, {and} \bibinfo{person}{Richard~P Eibach}.}
  \bibinfo{year}{2012}\natexlab{}.
\newblock \showarticletitle{Failing to commit: Maximizers avoid commitment in a
  way that contributes to reduced satisfaction}.
\newblock \bibinfo{journal}{{\em Personality and Individual Differences\/}}
  \bibinfo{volume}{52}, \bibinfo{number}{1} (\bibinfo{year}{2012}),
  \bibinfo{pages}{72--77}.
\newblock


\bibitem[\protect\citeauthoryear{Teppan and Felfernig}{Teppan and
  Felfernig}{2012}]%
        {teppan2012minimization}
\bibfield{author}{\bibinfo{person}{Erich~Christian Teppan} {and}
  \bibinfo{person}{Alexander Felfernig}.} \bibinfo{year}{2012}\natexlab{}.
\newblock \showarticletitle{Minimization of decoy effects in recommender result
  sets}.
\newblock \bibinfo{journal}{{\em Web Intelligence and Agent Systems: An
  International Journal\/}} \bibinfo{volume}{10}, \bibinfo{number}{4}
  (\bibinfo{year}{2012}), \bibinfo{pages}{385--395}.
\newblock


\bibitem[\protect\citeauthoryear{Tintarev and Masthof}{Tintarev and
  Masthof}{2015}]%
        {Tintarev2015ExplainingEvaluation}
\bibfield{author}{\bibinfo{person}{Nava Tintarev} {and} \bibinfo{person}{Judith
  Masthof}.} \bibinfo{year}{2015}\natexlab{}.
\newblock \showarticletitle{{Explaining recommendations: design and
  evaluation}}.
\newblock In \bibinfo{booktitle}{{\em Recommender Systems Handbook}}.
  \bibinfo{publisher}{Springer US}, \bibinfo{address}{Boston, MA},
  \bibinfo{pages}{217--253}.
\newblock
\showISBNx{978-1-4899-7637-6}
\showISSN{14779234}
\showDOI{%
\url{https://doi.org/10.1007/978-1-4899-7637-6}}


\bibitem[\protect\citeauthoryear{Wind, Green, Shifflet, and Scarbrough}{Wind
  et~al\mbox{.}}{1989}]%
        {wind1989courtyard}
\bibfield{author}{\bibinfo{person}{Jerry Wind}, \bibinfo{person}{Paul~E Green},
  \bibinfo{person}{Douglas Shifflet}, {and} \bibinfo{person}{Marsha
  Scarbrough}.} \bibinfo{year}{1989}\natexlab{}.
\newblock \showarticletitle{Courtyard by Marriott: Designing a hotel facility
  with consumer-based marketing models}.
\newblock \bibinfo{journal}{{\em Interfaces\/}} \bibinfo{volume}{19},
  \bibinfo{number}{1} (\bibinfo{year}{1989}), \bibinfo{pages}{25--47}.
\newblock


\bibitem[\protect\citeauthoryear{Yoo, Gretzel, and Zanker}{Yoo
  et~al\mbox{.}}{2012}]%
        {yoo2012persuasive}
\bibfield{author}{\bibinfo{person}{Kyung-Hyan Yoo}, \bibinfo{person}{Ulrike
  Gretzel}, {and} \bibinfo{person}{Markus Zanker}.}
  \bibinfo{year}{2012}\natexlab{}.
\newblock \bibinfo{booktitle}{{\em Persuasive recommender systems: conceptual
  background and implications}}.
\newblock \bibinfo{publisher}{Springer Science \& Business Media}.
\newblock


\bibitem[\protect\citeauthoryear{Zanker and Schoberegger}{Zanker and
  Schoberegger}{2014}]%
        {Zanker2014AnSystems}
\bibfield{author}{\bibinfo{person}{Markus Zanker} {and} \bibinfo{person}{Martin
  Schoberegger}.} \bibinfo{year}{2014}\natexlab{}.
\newblock \showarticletitle{{An empirical study on the persuasiveness of
  fact-based explanations for recommender systems}}. In
  \bibinfo{booktitle}{{\em CEUR Workshop Proceedings}},
  Vol.~\bibinfo{volume}{1253}. \bibinfo{pages}{33--36}.
\newblock
\showISSN{16130073}
\showURL{%
\url{http://ceur-ws.org/Vol-1253/paper6.pdf}}


\bibitem[\protect\citeauthoryear{Zwerina, Huber, and Kuhfeld}{Zwerina
  et~al\mbox{.}}{1996}]%
        {Zwerina1996ADesigns}
\bibfield{author}{\bibinfo{person}{Klaus Zwerina}, \bibinfo{person}{Joel
  Huber}, {and} \bibinfo{person}{Wf~Warren Kuhfeld}.}
  \bibinfo{year}{1996}\natexlab{}.
\newblock \showarticletitle{{A general method for constructing efficient choice
  designs}}.
\newblock \bibinfo{journal}{{\em Durham, NC: Fuqua School of Business, Duke
  Univesrsity\/}} \bibinfo{number}{September} (\bibinfo{year}{1996}),
  \bibinfo{pages}{39--59}.
\newblock
\showURL{%
\url{http://support.sas.com/techsup/technote/mr2010e.pdf}}


\end{thebibliography}

\end{document}